\documentclass[lettersize,journal]{IEEEtran}
\usepackage[T1]{fontenc}
\usepackage[utf8]{inputenc}
\usepackage{epsfig}
\usepackage{amsmath}
\usepackage{amssymb}
\usepackage{cite}
\usepackage{color}
\usepackage{url}
\usepackage{amsthm}
\usepackage{algorithm}
\usepackage{algorithmic}
\usepackage{multirow}
\usepackage{subfigure}
\usepackage{tabularx}
\usepackage{array}
\usepackage{varwidth}
\usepackage{setspace}
\usepackage{enumitem}
\usepackage{comment}
\usepackage{mathtools, cuted}
\usepackage{float}
\usepackage{stfloats}
\theoremstyle{proposition}

\theoremstyle{definition}
\theoremstyle{remark}
\newtheorem{remark}{Remark}
 
\def\log{{\mathrm{log}}}

\def\Re{{\mathfrak{Re}}}
\def\Im{{\mathfrak{Im}}}

\def\argmin{{\mathrm{argmin}}}
\def\argmax{{\mathrm{argmax}}}

\def\disp{\displaystyle}

\def\QCS-CE{{\mathrm{QCS}\text{-}\mathrm{CE}}}

\newcommand{\veclatin}[1]{{\bf{#1}}}

\def\ab{{\veclatin{a}}}

\def\Ab{{\veclatin{A}}}

\def\Gb{{\veclatin{G}}}
\def\hb{{\veclatin{h}}}

\def\Hb{{\veclatin{H}}}

\def\nb{{\veclatin{n}}}

\def\rb{{\veclatin{r}}}

\def\xb{{\veclatin{x}}}

\def\yb{{\veclatin{y}}}

\def\zb{{\veclatin{z}}}

\begin{document}

\title{Regularized Neural Detection for One-Bit Massive MIMO Communication Systems}

\author{Aditya~Sant,~\IEEEmembership{Graduate Student Member,~IEEE,}
and Bhaskar~D.~Rao,~\IEEEmembership{Fellow,~IEEE}
\thanks{This work is financially supported in part by ONR Grant No. N00014-18-1-2038, NSF grant CCF-2124929, NSF Grant CCF-2225617, and the UCSD Center for Wireless Communications.} 
}



\maketitle

\begin{abstract}
Detection for one-bit massive MIMO systems presents several challenges especially for higher order constellations.
Recent advances in both model-based analysis and deep learning frameworks have resulted in several robust one-bit detector designs.
Our work builds on the current state-of-the-art gradient descent (GD)-based detector. We introduce two novel contributions in our detector design: \textit{(i)} We augment each GD iteration with a deep learning-aided regularization step, and \textit{(ii)} We introduce a novel constellation-based loss function for our regularized DNN detector. 
This one-bit detection strategy is applied to two different DNN architectures based on algorithm unrolling, namely, a deep unfolded neural network and a deep recurrent neural network. Being trained on multiple randomly sampled channel matrices, these networks are developed as general one-bit detectors.
%Further, the general regularized GD and robust constellation aware loss function can be potentially extended to multi-bit received data.
The numerical results show that the combination of the DNN-augmented regularized GD and constellation-based loss function improve  the quality of our one-bit detector, especially for higher order M-QAM constellations.
\end{abstract}


\section{Introduction}\label{Sec_Intro}
\noindent \IEEEPARstart{N}{ext} generation massive MIMO communication system design promises high-speed wireless communication and an entire network of interconnected devices \cite{bana2019massive,shafique2020internet}. However,  widescale deployment brings in challenges for system cost, power consumption and complexity. Several advances in model-based algorithm design as well as high performance DNNs are being made to combat these challenges for both channel estimation as well as end-to-end communication. The general parametric structure of DNNs, coupled with their advantage as universal functional approximators \cite{hornik1989multilayer,goodfellow2016deep}, makes these an integral part of the future of robust wireless communication, exploited for a variety of applications from beamformer design \cite{sohrabi2021deep,zhang2021reinforcement,sant2022deep}, channel estimation \cite{ye2017power,wen2018deep,soltani2019deep} as well as end-to-end detection \cite{o2017deep,diskin2017deep,he2020model,khani2020adaptive,pratik2020re}.

One of the major challenges for widescale deployment is the design of high-resolution analog to digital converters (ADCs). Prior analysis of system design has shown that high-resolution ADCs account for significant system cost and power consumption \cite{mo2017hybrid,murmann2020adc}.
Moving in the direction of low cost and complexity, low-resolution ADCs have been gaining significant interest, due to advances in both signal processing and DNN-based algorithms \cite{nguyen2022deep,nguyen2020neural,mo2017channel,mo2017hybrid,fan2015uplink,jacobsson2017throughput}. A special case of low-resolution ADCs is the one-bit ADC. One-bit signal recovery has seen various innovations in general signal processing research \cite{li2018survey,knill1998power,boufounos20081,khobahi2019deep}. In our work, we focus on the application of DNN-based methods to  symbol recovery for one-bit massive MIMO communication systems. DNN based detectors appear to be naturally suited for this problem because of the inherent nonlinearity in the measurement process.

One-bit MIMO data detection benefited significantly with the application of Bussgang's theroem to linearize the input-output relation \cite{bussgang1952crosscorrelation}. Based on this linearization, a large class of linear receivers as well as MMSE receivers has been proposed for both single carrier and multi-carrier systems \cite{mezghani2007modified,ho2019antithetic,wan2020generalized}. In addition several works utilize this linearization to characterize the one-bit system and evaluate the overall system performance and capacity \cite{jacobsson2015one,mollen2016uplink,li2017channel}. Additional robust model-based detectors improving on the Bussgang linear detectors have also been proposed in several key works \cite{nguyen2021svm,thoota2021variational}. In addition to one-bit data detection, one bit channel estimation for mmWave communication systems has also been studied \cite{stockle2016channel,liu2017one}. Our previous work \cite{sant2020doa} characterizes the subspace of the one-bit transformed signal and even generalizes this behavior to a broader class of odd-symmetric nonlinearities. In addition to the different model-based approaches, different works applying DNNs to one-bit detection have also been proposed \cite{balevi2019one,kim2019semi,nguyen2021dnn,khobahi2021lord,zhang2020deep,kim2020machine,jeon2022artificial}. 

One of the most resilient class of one-bit detectors is based on the one-bit likelihood maximization of the received signal using the Gaussian cumulative distribution function (cdf) \cite{choi2015quantized}. The work in \cite{choi2016near} introduced a near maximum likelihood (n-ML) detector based on a two step iterative algorithm - gradient descent (GD) followed by projection onto the unit sphere. Other works applying the Gaussian cdf likelihood formulation have also been used extending this idea \cite{jeon2018one,jeon2019robust}. However, one of the limitations of applying the GD iteration on the Gaussian cdf is its instability at high signal-to-noise ratio (SNR) values \cite{Ho2022Thesis}. The work in \cite{nguyen2021linear}
applied the sigmoid approximation of the Gaussian cdf \cite{bowling2009logistic} to the one-bit likelihood. The ensuing detector, that the authors named the OBMNet, formulated this detection as an unfolded DNN, learning the GD step sizes at each iteration. The sigmoid approximation was shown empirically to stabilize the gradient and addressed more explicitly in \cite{Ho2022Thesis}. This work is, at present, the current state-of-the-art for M-QAM data detection.

Our work builds on the state-of-the-art OBMNet formulation, with the following contributions.
\begin{itemize}
    \item We introduce a novel, regularized GD approach for one bit detection. We augment each GD iteration with a learnable DNN-based step. This DNN-based step performs an explicit regularization of each GD iteration of the OBMNet algorithm, enhancing recovery for data symbols transmitted from an M-QAM constellation. We capitalize on not only the model-based OBMNet structure, but also increase the network expressivity through this DNN-aided regularization block per iteration.
    \item We improve on the generalization capability of existing end-to-end detection networks (mentioned earlier), which are trained and tested on a single channel response. By designing the architecture, input data as well as training on on multiple randomly sampled Rayleigh-fading channels, we avoid the need to re-train the detector network for each different channel state information matrix.
    \item We implement two unique networks, for the above mentioned regularized GD approach: 
    \begin{enumerate}
        \item ROBNet: A deep unfolded network with an identical, \textit{different}, sub-network block per GD iteration
        \item OBiRIM: A deep recurrent neural network utilizing estimation memory for one-bit estimation
    \end{enumerate}
    To the best of our knowledge, the latter, i.e. the OBiRIM, presents the first approach using a recurrent neural network for one-bit detection. 
    \item Contrary to the mean square error (MSE) loss used for network training, we introduce a novel loss function, tailored to MIMO communication symbol recovery. In particular, we incorporate a constellation-aware regularized MSE loss function to penalize the symbol errors as well as the bit errors. We envision this as a general communication system loss function, not just limited to one-bit symbol recovery.
\end{itemize}
Our experimental results, implemented on the i.i.d. Rayleigh fading channel, show the utility of considering a robust regularized GD algorithm through sharper and more compact recovered constellation clusters with significantly reduced cluster spread. This improved recovery is especially significant for improved detection performance of higher order M-QAM constellations. Although the analysis of multi-bit MIMO receivers falls outside the scope of this work, the presented regularized GD framework and robust constellation aware DNN loss function can potentially be applied to deal with the nonlinearities of these systems as well.

%Although we have introduced a novel DNN-aided regularization to the one-bit MIMO framework, the explicit learnable quantization-based projection in \cite{nguyen2022deep} is shown to outperform our proposed framework. We have provided relevant experimental validation and reasoning in Sec. \ref{Sec_exp_results}.
The purpose of this document is provide background and details necessary for the mmWave extension presented at IEEE ICASSP 2023.

\smallskip

\noindent \textit{Organization:} This manuscript is organized as follows - Sec. \ref{Sec_sys_model_background} introduces the system model, one-bit detection problem and the gradient-descent based approaches used. Sec. \ref{Sec_reg_one_bit_detector} introduces our proposed framework for general regularized one-bit detection, while Sec. \ref{Sec_reg_gd_dnn_implementation} explains the specific DNN implementation used. Sec. \ref{Sec_exp_results} provides experimental validation of our proposed framework and Sec. \ref{Sec_conclusions} concludes the manuscript.

\smallskip

\noindent \textit{Notation:} We use lower-case boldface letters $\ab$ and upper case boldface letters $\Ab$ to denote complex valued vectors and matrices respectively. The notation $\Re(\cdot)$ and $\Im(\cdot)$ denote the real and imaginary parts, respectively. The operation $(\cdot)^{\mathrm{T}}$ denotes the transpose of the array or matrix. Unless otherwise specified, all scalar functions like $\mathrm{tanh}(\cdot)$ or $\mathrm{sign}(\cdot)$, when applied to arrays or matrices, imply element-wise operation.
The notation $\xb^{(t)}$ is used to denote the value of the variable $\xb$ at iteration $t$ of the algorithm. For the DNN training, the size of the training set is given by $N_{\mathrm{train}}$ and the notation $\hat{\xb}_{n,\mathrm{train}}$ denotes the $n^{\mathrm{th}}$ sample from this set. Unless otherwise specified, the norm $||\cdot||$ represents the $\ell_{2}$-norm for a vector and Frobenius norm for a matrix.

\section{System Model and Background}\label{Sec_sys_model_background}
\noindent In this section we introduce the wireless system model, followed by the one-bit maximum likelihood (ML) optimization, {resulting in the GD-based detector}.
Finally, we review the OBMNet framework, which forms an integral part of our network structure. We specifically address the strengths and observed shortcomings that we address through our work.

\subsection{System model}
\noindent We utilize the same random channel with block flat-fading  as in most past works, e.g. \cite{tse2005fundamentals,barsocchi2006channel}. We consider $K$ single antenna users transmitting to a multi-antenna base-station (BS) with $N$ receive antennas. The MIMO channel $\bar{\Hb}\in\mathbb{C}^{N\times K}$ consists of i.i.d {entries drawn from $\mathcal{CN}(0,1)$}. We assume the BS has perfect unquantized channel state information (CSI). However, through our experimental results in Sec. \ref{Sec_exp_results}, we also model imperfect the CSI to the detector.

As a part of the multi-user uplink, the $k^{\mathrm{th}}$ user transmits the signal $\bar{x}_{k}$, drawn from the M-QAM constellation. The overall transmitted signal is $\bar{\xb}=\begin{bmatrix}\bar{x}_{1},\bar{x}_{2},\hdots,\bar{x}_{K}\end{bmatrix}^{\mathrm{T}}$. The unquantized received signal at the BS is given by 
\begin{equation}\label{eq_unqunat_sig}
    \bar{\rb}=\bar{\Hb}\bar{\xb} + \bar{\zb},
\end{equation}
where $\bar{\zb}$ is the AWGN with noise variance depending on the system signal-to-noise ratio (SNR) $\rho=\frac{\mathbb{E}(||\bar{\Hb}\bar{\xb}||^{2})}{\mathbb{E}(||\bar{\zb}||^{2})}$. The transformed signal due to the one-bit quantization is given by 
\begin{equation}\label{eq_one_bit_quant}
    \bar{\yb}=\mathrm{sign}\big{(}\Re(\bar{\rb})\big{)}+j\,\mathrm{sign}\big{(}\Im(\bar{\rb})\big{)}.
\end{equation}

In order to express the algorithm design as a function of  real-valued inputs, we convert the received signal and the observed channel matrix into real-valued forms as
\begin{equation}\label{eq_convert_real_value}
    \begin{gathered}
    \Hb=\begin{bmatrix}\Re(\bar{\Hb}) & -\Im(\bar{\Hb})\\
    \Im(\bar{\Hb})&\Re(\bar{\Hb})\end{bmatrix}, \
    \xb=\begin{bmatrix}\Re(\bar{\xb}) \\ \Im(\bar{\xb})\end{bmatrix}, \\
    \rb=\begin{bmatrix}\Re(\bar{\rb}) \\ \Im(\bar{\rb})\end{bmatrix}, \
    \yb=\begin{bmatrix}\Re(\bar{\yb}) \\ \Im(\bar{\yb})\end{bmatrix}, \
    \zb=\begin{bmatrix}\Re(\bar{\zb}) \\ \Im(\bar{\zb})\end{bmatrix}.
    \end{gathered}
\end{equation}
Thus, the modified received one-bit signal at the BS is
\begin{equation}
    \yb=\mathrm{sign}(\Hb \xb + \zb).
\end{equation}
{The detection algorithm recovers the transmitted symbols $\xb$ from the one-bit received data $\yb$}.

\subsection{{One-bit maximum likelihood and GD-based detection}}
\noindent The one-bit maximum likelihood (ML) problem has been derived in \cite{choi2015quantized} as
\begin{equation}\label{eq_one_bit_ml_prob}
        \hat{\xb}_{\mathrm{ML}}=\disp\underset{\xb\in\mathcal{M}^{2K}}{\argmax} \sum_{i=1}^{2N}\,\log \, \Phi\big{(}\sqrt{2\rho}\,y_{i}\hb_{i}^{\mathrm{T}}\xb\big{)},
\end{equation}
where $\Phi(\cdot)$ is the cumulative distribution function (cdf) for $\mathcal{N}(0,1)$ and $\mathcal{M}^{2K}$ represents the set of the $2K$-dimensional vectors, consisting of the real-valued representation (see eq. \eqref{eq_convert_real_value}) of the $K$-dimensional vectors of M-QAM constellation symbols. The search over this constrained, finite, non-convex set $\mathcal{M}^{2K}$ scales this problem exponentially in the number of users. Different approaches based on relaxations of the optimization \eqref{eq_one_bit_ml_prob} have been proposed \cite{choi2016near,nguyen2021linear,khobahi2021lord}.

One of the proposed relaxations for the constrained optimization \eqref{eq_one_bit_ml_prob} involves unconstrained GD over the entire subspace $\mathbb{R}^{2K}$, followed by a projection onto the subspace of interest \cite{choi2016near}. The unconstrained GD update step has been derived in \cite{choi2016near} as
\begin{equation}
        \xb^{(t+1)}=\xb^{(t)} + \alpha^{(t)}\sqrt{2\rho}\,\Gb^{\mathrm{T}}\frac{\phi(\sqrt{2\rho}\,\Gb\xb)}{\Phi(\sqrt{2\rho}\,\Gb\xb)}
\end{equation}
where $\alpha^{(t)}$ is the step size at iteration $t$, $\Gb=\mathrm{diag}(y_{1},y_{2},\hdots,y_{2N})\,\Hb$ and $\phi(\cdot)$ is the Gaussian probability density function. The subsequent step projects this estimate $\xb^{(t+1)}$ onto the unit hyper-sphere.

This optimization approach is limited by the behavior of the Gaussian cdf $\Phi(\cdot)$ at high SNR values. It is empirically observed that this function drops rapidly to zero at high SNR values, making the likelihood gradient explode to large values. Further the Hessian matrix for the same is empirically observed to contain a high condition number \cite{Ho2022Thesis}. All this makes the optimization \eqref{eq_one_bit_ml_prob} unstable at high SNR values.

\subsection{Current state-of-the-art one-bit detector: OBMNet}\label{Sec_obmnet}
\noindent 
An approximate ML estimation framework was proposed in \cite{nguyen2021linear} using the logistic cdf  approximation of the Gaussian cdf \cite{bowling2009logistic}. This approximation involves sigmoids, a popular activation function in neural networks, and naturally leads to a DNN based detector. The authors in \cite{nguyen2021linear} empirically observe a robustness in detection to incorrect symbol estimation as well as imperfect CSI at the detector as a result of this approximation. This can be explained by examining the gradient of the approximate ML and noting that it is much better behaved at high SNR \cite{Ho2022Thesis}.
The approximate ML problem using the sigmoid log-likelihood is given by
\begin{equation}\label{eq_one_bit_ml_sigmoid}
    \hat{\xb}_{\mathrm{ML}}
        =\disp\underset{\xb\in\mathcal{M}^{2K}}{\argmin} \sum_{i=1}^{2N}\,\log \, (1+e^{-c\sqrt{2\rho}y_{i}\hb_{i}^{\mathrm{T}}\xb}),
\end{equation}
with the value of $c=1.702$. {Applying GD to the likelihood \eqref{eq_one_bit_ml_sigmoid}, we have the update equation}
\begin{equation}\label{eq_obmnet_gd_update}
    \begin{aligned}
        \xb^{(t+1)}&=\xb^{(t)}-\alpha^{(t)}\nabla^{(t)}_{\xb}\\
        &=\xb^{(t)} + \alpha^{(t)}\Gb^{\mathrm{T}}\sigma(-\Gb\xb^{(t)}), \ \  t=0,\hdots,T-1,
    \end{aligned}
\end{equation}
where $\sigma(\cdot)$ is the logistic sigmoid function. The constants have been absorbed into the step size $\alpha^{(t)}$. {After executing $T$ iterations of GD, the final estimate $\hat{\xb}^{(T)}$ is normalized as}
\begin{equation}\label{eq_obmnet_normalize}
    \tilde{\xb}=\frac{\sqrt{K}}{||\xb^{(T)}||}\hat{\xb}^{(T)}.
\end{equation}
{The $T$-step unconstrained update \eqref{eq_obmnet_gd_update} is implemented as a $T$-layer unfolded DNN with sigmoid nonlinearity and network weights depending on the CSI matrix and one-bit measurements, i.e., the OBMNet \cite{nguyen2021linear}. The step sizes at each iteration $\alpha^{(t)}$ are the only learnable parameters.} The network parameters are trained on the MSE loss function 
\begin{equation}\label{eq_obmnet_loss_func}
    \mathcal{L}=\frac{1}{N_{\mathrm{train}}}\sum_{n=1}^{N_\mathrm{train}}||\tilde{\xb}_{n}-\tilde{\xb}_{\mathrm{train},n}||^{2}.
\end{equation}
The results in \cite{nguyen2021linear} show the OBMNet as an efficient low-complexity detector for QPSK as well as 16-QAM symbols. However this detector has a few limitations, described below.
\begin{enumerate}[label=(\roman*)]
    \item \textit{Limited network expressivity:} The OBMNet, as a general DNN, is highly underparameterized. Any changes in the network architecture, loss function and training procedure do not show up in improved performance for the network.
    \item \textit{Constellation cluster spread:} Analyzing scatter plots for the recovered symbols, high cluster spread is evident (see Fig.~\ref{fig_const_recovered_qpsk}). More on this in the next section. Although this does not compromise bit error rate (BER) for lower order constellations like QPSK, it degrades performance at higher order constellations like 16-QAM.
    \item \textit{Gap to ML:} The original two-step OBMNet detection \eqref{eq_obmnet_gd_update}-\eqref{eq_obmnet_normalize} falls short of the theoretical exponential search based ML solution to \eqref{eq_one_bit_ml_sigmoid}. The authors in \cite{nguyen2021linear} fine-tune their estimates through a constrained lower order ML search step to bridge this gap.
\end{enumerate}
%
%
%\textit{(i)} Limited network expressivity: The OBMNet, as a general DNN, is highly underparameterized. Any changes in the network architecture, loss function and training procedure do not show up in improved performance for the network. \textit{(ii)} Constellation cluster spread: On observing the constellations for the recovered symbols, a high cluster spread is evident. Although this does not compromise bit error rate (BER) performance for lower order constellations like QPSK, it degrades performance at higher order constellations like 16-QAM. \textit{(iii)} Gap to ML: The original two-step OBMNet detection \eqref{eq_obmnet_gd_update}-\eqref{eq_obmnet_normalize} falls short of the theoretical exponential search based ML solution to \eqref{eq_one_bit_ml_sigmoid}. The authors in \cite{nguyen2021linear} fine-tune their estimates through a constrained lower order ML search step to bridge this gap.
%The OBMNet thus still has significant potential for improved one-bit detector design.

\section{Regularized Neural Detection Framework}\label{Sec_reg_one_bit_detector}
\noindent In order to address some of the observed limitations of the OBMNet, we introduce the framework of regularized neural one-bit detection, building on the OBMNet framework (Sec. \ref{Sec_obmnet}).
The specific network structure and implementation details for our approach are provided in the next section. Here, we begin with the general regularized GD framework, with a learnable DNN-aided regularization.
Next, for robust DNN training, we have developed a novel constellation-aware quantization based loss function.
Finally we comment on the ability to generalize to any arbitrary Rayleigh fading channel.

\subsection{DNN-aided regularized GD for one-bit MIMO detection}
\noindent In order to improve the detection robustness, we modify the unconstrained OBMNet update step \eqref{eq_obmnet_gd_update} to a regularized GD update, per iteration $t$, given by 
{
\begin{subequations}\label{eq_reg_one_bit_update}
    \begin{align}
        \hat{\xb}^{(t+1)} &= \xb^{(t)} - \alpha^{(t)}\nabla_{\xb}^{(t)}\label{eq_reg_one_bit_update_gd_update} \\
        \xb^{(t+1)} &= \hat{\xb}^{(t+1)} + h_{\phi}^{(t)}(\xb^{(t)},\nabla_{\xb}^{(t)},\hat{\xb}^{(t+1)})\label{eq_reg_one_bit_update_reg_gd_update}.
    \end{align}
\end{subequations}
}
Here, the first step {\eqref{eq_reg_one_bit_update_gd_update}}, the intermediate update, is the same as the unconstrained OBMNet update \eqref{eq_obmnet_gd_update}. 
{The second step \eqref{eq_reg_one_bit_update_reg_gd_update} represents the introduced correction to this unconstrained update. Based on regularizing the estimate $\hat{\xb}^{(t+1)}$ to account for optimization within the M-QAM constellation space, the overall update \eqref{eq_reg_one_bit_update} is called the regularized GD detection for one-bit MIMO.}
We introduce a parametric regularization function $h_{\phi}^{(t)}(\cdot)$, per iteration, implemented via a DNN (exact implementation in Sec. \ref{Sec_reg_gd_dnn_implementation}). By means of an additional learnable regularization we increase the network expressivity of the original OBMNet, by increasing the number of learnable network parameters. We also enable per-iteration projection of the iterand $\xb^{(t)}$ onto the set of the real-valued representation of M-QAM constellation points $\mathcal{M}^{2K}$.
%As a consequence of this learnable regularization, as shown later, we enable symbol recovery with very low cluster spread.
%In order to capitalize on this general parametric regularization structure, we design a constellation-aware loss function. This is further elaborated in the subsequent subsection. 

{
The detector FBM-DetNet, introduced in \cite{nguyen2022deep}, also implements a per-iteration projection of the OBMNet estimate on the $\mathcal{M}^{2K}$ subspace at each iteration. This is implemented using a learnable hard quantization of each iterand $\hat{\xb}^{(t+1)}$ to the M-QAM constellation. Differently, the regularized GD \eqref{eq_reg_one_bit_update} learns a general projection function, implemented as a residual correction at each step.
}

\subsection{Improved DNN loss function}\label{Sec_improved_loss}
\noindent 
In order to capitalize on the general parametric regularization structure, we design a constellation-aware loss function.
The MSE loss function \eqref{eq_obmnet_loss_func}, utilized by the OBMNet, penalizes the magnitude of the symbol error for the received signal. We attempt to add in an additional robustness to network training by also penalizing symbol flips in the estimated symbols, thus implicitly penalizing bit flips in the recovered data.

\begin{figure}[t]
	\centering
	\includegraphics[width=\linewidth]{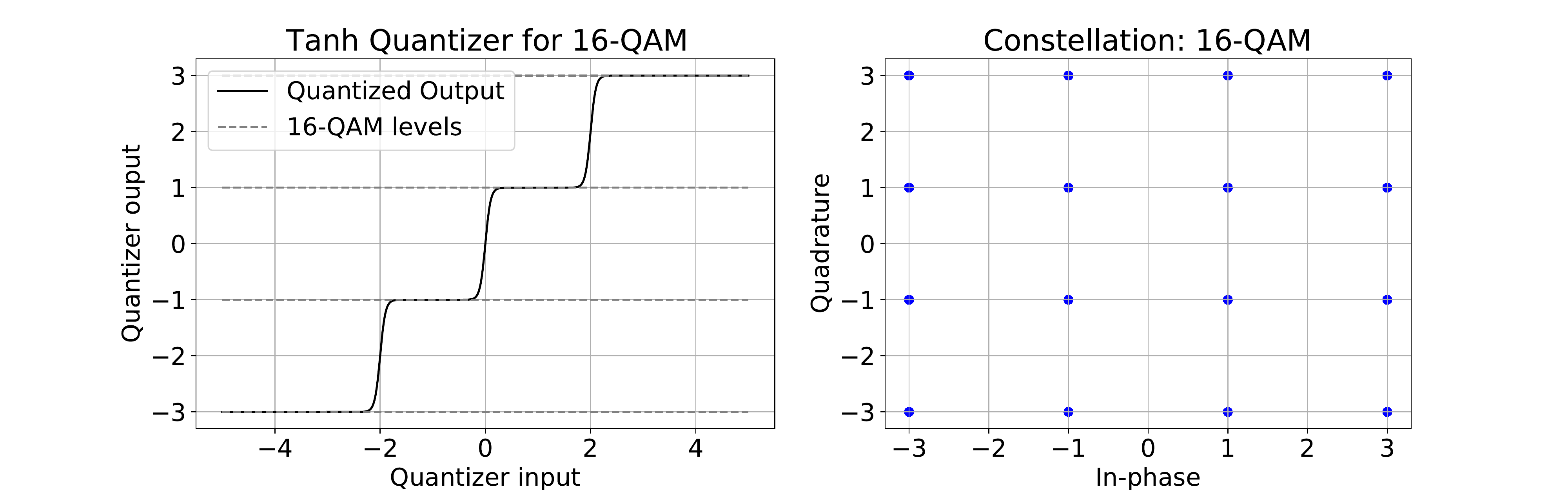}
	\caption{Illustration for 16-QAM quantizer \eqref{eq_quant_16qam}. The value of $\beta=10$.}
	\label{fig_tanh_quantizer_16_qam}
\end{figure}

Incorporating this robustness, we improve on the MSE loss by using the following modification
\begin{equation}\label{eq_reg_one_bit_loss_func}
    \mathcal{L}=\frac{1}{N_{\mathrm{train}}}\sum_{n=1}^{N_{\mathrm{train}}}\,\big{[}||\xb^{(T)}_{n}-\tilde{\xb}_{\mathrm{train},n}||^{2}+\lambda\, \mathcal{R}(\xb^{(T)}_{n},\tilde{\xb}_{\mathrm{train},n})\big{]},
\end{equation}
where $\mathcal{R}(\cdot)$ is a constellation-aware regularization for DNN training. This regularization is based on a smooth quantization of the network output, and implemented as
\begin{equation}\label{eq_reg_one_bit_quant_loss}
    \mathcal{R}(\xb^{(T)}_{n},\tilde{\xb}_{\mathrm{train},n})=||\mathcal{Q}_{\beta}(\xb^{(T)}_{n})-\tilde{\xb}_{\mathrm{train},n}||^{2}.
\end{equation}
Here, the function $\mathcal{Q}_{\beta}(\cdot)$ is a smooth constellation-aware quantization function, utilizing the nonlinearity $f_{\beta}(z) = \mathrm{tanh}(\beta z)$ with a hyperparameter $\beta$. {The choice of the scaled $\mathrm{tanh}(\cdot)$ nonlinearity is inspired by \textit{(i)} The saturating behavior for quantization, \textit{(ii)} Differentiability for backpropagation of the loss, and \textit{(iii)} Ease of tuning to regulate the quantization degree. For the two considered constellations in this work, we implement the quantization function $\mathcal{Q}_{\beta}(\cdot)$ as follows.}
\begin{enumerate}
    \item $\mathcal{Q}_{\beta}(x)$ for QPSK constellation:
    \begin{equation}\label{eq_quant_qpsk}
        \mathcal{Q}_{\beta}(x)=\mathrm{tanh}(\beta\,x).
    \end{equation}
    \item $\mathcal{Q}_{\beta}(x)$ for 16-QAM constellation:
    \begin{equation}\label{eq_quant_16qam}
        \mathcal{Q}_{\beta}(x)=\mathrm{tanh}(\beta\,(x+2))+\mathrm{tanh}(\beta\,x)+\mathrm{tanh}(\beta\,(x-2)).
    \end{equation}
\end{enumerate}
The quantization function \eqref{eq_quant_16qam} for the 16-QAM constellation is plotted in Fig. \ref{fig_tanh_quantizer_16_qam}. The plots illustrate that the quantization \eqref{eq_quant_16qam} implements a smooth version of the symbol-mapper, that can be backpropogated through the regularization network, to the 16-QAM constellation symbols.
By specifically modifying the loss function as \eqref{eq_reg_one_bit_loss_func} to communication system symbol recovery, we are able to incorporate a symbol error rate (SER) metric into the training phase of our networks. The role of the quantizer \eqref{eq_reg_one_bit_quant_loss} is to cluster the estimated symbols within a very small neighborhood of the nearest M-QAM symbols. Thus the symbol loss for the staying within the ``right'' symbol boundaries is attenuated and the symbol loss for crossing over the symbol boundaries is amplified. Thus the regularization loss will be dominated by symbol errors (implicitly bit errors). By incorporating this into the training phase, we also account for an improved BER performance, a metric that is imperative to communication system design.
\begin{remark}\label{remark_improved_dnn_loss}
    {The general quantization function $\mathcal{Q}_{\beta}(\cdot)$ can be implemented using different nonlinearites, like the ReLU. Analysis of such alternate quantization functions will extract larger patterns in the behavior of the constellation-aware regularization. This work conceptually introduces improved loss functions through the use of M-QAM constellation mapping, in order to incorporate symbol error rate in the network training. The specific detailed analysis of alternate loss functions falls outside the scope of this work.}
\end{remark}

Before illustrating our specific implementations of this neural detector, we present another advantage of this framework, i.e., the generalizability to arbitrary Rayleigh fading channels.

\subsection{Generalization of one-bit neural detection}\label{Sec_gen_detect}
\noindent Contrary to the conventional end-to-end learning approaches for one-bit detection \cite{balevi2019one,kim2019semi,khobahi2021lord}, we develop our regularized model beyond a channel-specific detector. The regularization network $h_{\phi}^{(t)}(\cdot)$ in \eqref{eq_reg_one_bit_update} implicitly takes in the channel information via the gradient of the estimate at each iteration $\nabla_{\xb}^{(t)}$, which is used for the signal recovery. Each new subsequent channel matrix $\Hb$ results in a new sequence of gradient expressions for the unfolded network, i.e., $\nabla_{\xb}^{(t)}$. This, in turn, enables the network to uniquely identify the inputs with the channel response. As opposed to directly feeding in the input channel matrix $\Hb$ to the regularization network, our approach exploits the main advantage of unfolded deep learning \cite{monga2021algorithm} by using the channel information in an appropriate, model-based form. By feeding the gradient of the signal, for any generated channel matrix $\Hb$, to the regularization network, we are able to efficiently fine tune the original GD algorithm for that particular channel matrix $\Hb$. 
This approach of learning a parametric regularization from the gradient of the linear model was also used for recurrent inference machines (RIMs) \cite{putzky2017recurrent}.
We thus overcome the need to re-train or fine-tune the network for each unique channel matrix $\Hb$. This enables completely eliminating the need to transmit any other additional pilot symbols (for any online training) following the initial access and channel state information (CSI) estimation phase.

\section{DNN-Aided Regularized GD: Implementation}\label{Sec_reg_gd_dnn_implementation}
\noindent Based on the proposed regularized GD and loss function framework introduced earlier, the next two subsections present the specific implementation via two distinct approaches, namely, the unfolded ROBNet and the recurrent OBiRIM.

\subsection{Unfolded one-bit DNN: ROBNet}\label{Sec_robnet}
\begin{figure}[t]
	\centering
	\includegraphics[width=\linewidth]{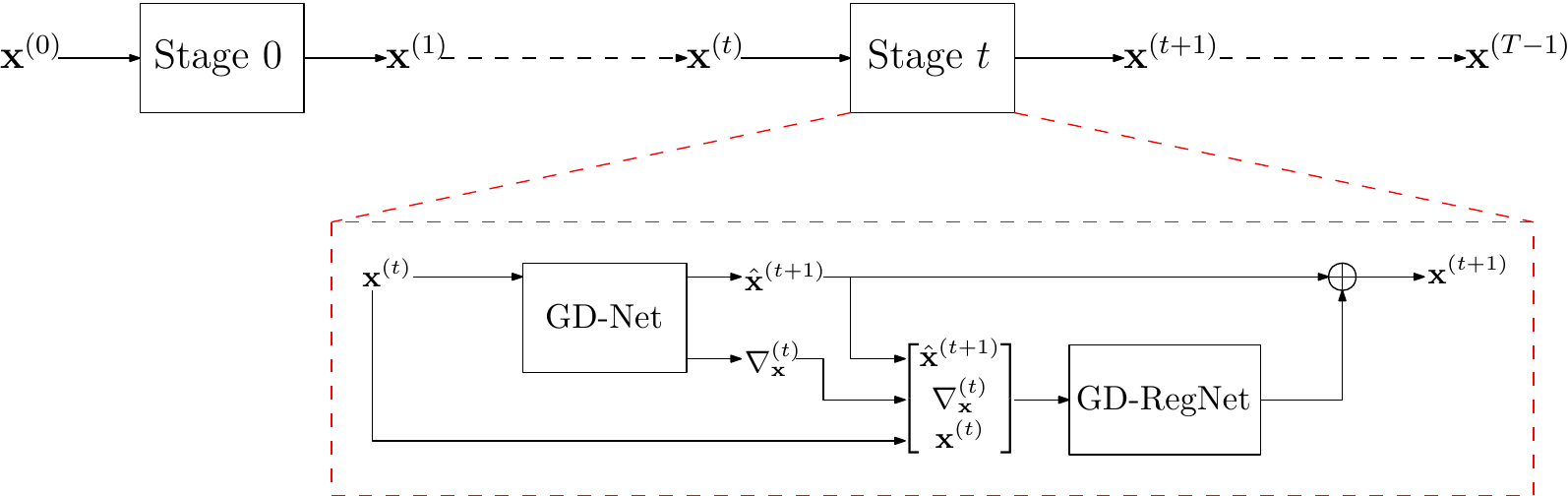}
	\caption{{Block diagram for the Regularized One-bit Detector (ROBNet)}}
	\label{fig_proj_robnet_overall}
\end{figure}
\begin{table}[t]
\caption{DNN Parameters {of GD-RegNet in} ROBNet \& OBiRIM (K Users)}
\label{table_dnn_param}
\centering
\begin{small}
\begin{tabular}{|l|l|l|}
\hline
Network                                                                         & Layer                                                                                 & Parameters                     \\ \hline
\multirow{11}{*}{\begin{tabular}[c]{@{}l@{}}ROBNet\\ (each stage)\end{tabular}} & \multirow{5}{*}{\textit{\begin{tabular}[c]{@{}l@{}}Convolution\\ (1-D)\end{tabular}}} & conv + ReLU + bn               \\ \cline{3-3} 
                                                                                &                                                                                       & Input dim - K                  \\ \cline{3-3} 
                                                                                &                                                                                       & Input chan - 6                 \\ \cline{3-3} 
                                                                                &                                                                                       & Output chan - 64               \\ \cline{3-3} 
                                                                                &                                                                                       & Kernel size - 3                \\ \cline{2-3} 
                                                                                & \multirow{6}{*}{\textit{Fully-connected}}                                             & Input dim - 64K                \\ \cline{3-3} 
                                                                                &                                                                                       & Output dim - 2K                \\ \cline{3-3} 
                                                                                &                                                                                       & Hidden layers - 3              \\ \cline{3-3} 
                                                                                &                                                                                       & Hidden dim - \{128,64,32\}     \\ \cline{3-3} 
                                                                                &                                                                                       & bn dim - \{128,64,32\}         \\ \cline{3-3} 
                                                                                &                                                                                       & Nonlinearity - ReLU            \\ \hline
\multirow{12}{*}{\begin{tabular}[c]{@{}l@{}}OBiRIM\\ (each stage)\end{tabular}} & \textit{Convolution}                                                                  & Same as ROBNet                 \\ \cline{2-3} 
                                                                                & \multirow{5}{*}{\textit{GRU}}                                                         & Num of GRUs - 2                \\ \cline{3-3} 
                                                                                &                                                                                       & GRU1 input dim - 64K           \\ \cline{3-3} 
                                                                                &                                                                                       & GRU1 hidden dim - 1024         \\ \cline{3-3} 
                                                                                &                                                                                       & GRU2 input dim - 1024          \\ \cline{3-3} 
                                                                                &                                                                                       & GRU2 hidden dim - 1024         \\ \cline{2-3} 
                                                                                & \multirow{6}{*}{\textit{Fully-connected}}                                             & Input dim - 1024               \\ \cline{3-3} 
                                                                                &                                                                                       & Output dim - 2K                \\ \cline{3-3} 
                                                                                &                                                                                       & Hidden layers - 4              \\ \cline{3-3} 
                                                                                &                                                                                       & Hidden dim - \{512,128,64,32\} \\ \cline{3-3} 
                                                                                &                                                                                       & bn dim - \{512,128,64,32\}     \\ \cline{3-3} 
                                                                                &                                                                                       & Nonlinearity - ReLU            \\ \hline
\end{tabular}
\end{small}
\end{table}
\noindent Model-based algorithm unrolling and the use of unfolded DNNs have been explored in different applications of signal processing and wireless communication \cite{balatsoukas2019deep,monga2021algorithm}. These networks are able to account for any model mismatch and can significantly save on the number of iterations, compared to the original model-based algorithms. The ability to use such network structures to complement model-based analysis motivates us to incorporate such an unfolded DNN to implement our regularized one-bit GD approach \eqref{eq_reg_one_bit_update}. 

Our proposed unfolded network implementation, the regularized one-bit network (ROBNet) is illustrated in Fig. \ref{fig_proj_robnet_overall}. Based on this, we present the following salient features of the unfolded learning approach.
\begin{itemize}
    \item The ROBNet, implementing a $T$-stage regularized GD algorithm, is unfolded into $T$ distinct sub-networks (each represented as Stage $t$ in Fig. \ref{fig_proj_robnet_overall}). {Each sub-network at Stage $t$ consists of two sequential phases.}
        \begin{enumerate}
            \item {GD-Net - Identical to each OBMNet \cite{nguyen2021linear} iteration, this implements \eqref{eq_reg_one_bit_update_gd_update}, with the $t^{\mathrm{th}}$ gradient and unconstrained iterate given by $\mathbf{\nabla}_{\xb}^{(t)}$ and $\hat{\xb}^{(t+1)}$, respectively. The GD step size $\alpha^{(t)}$ is the only learnable parameter.}
            \item {GD-RegNet - Denoted by $h_{\phi}^{(t)}(\hat{\xb}^{(t+1)},\mathbf{\nabla}_{\xb}^{(t)},\xb^{(t)})$, this is a larger parametric network that regularizes each GD iteration, i.e., \eqref{eq_reg_one_bit_update_reg_gd_update}. This increases network exressivity through a larger number of learnable parameters.}
        \end{enumerate}
    \item Additionally, for each Stage $t$, a residual link from the {GD-Net output $\hat{\xb}^{(t)}$} is fed to the output of the {GD-RegNet}. Thus the role of each {GD-RegNet} at Stage $t$ is to impart an appropriate stage-dependent correction, learnt from the data, to the unconstrained gradient step. 
\end{itemize}
We now provide the specific technical details {of this GD regularization}, along with the {general channel training}.

\medskip
\noindent \textit{{GD-RegNet} structure and training:}
We begin by describing the input to the {GD-RegNet} at each Stage $t$, consisting of the {GD-Net output - } the unconstrained update $\hat{\xb}^{(t+1)}$, gradient $\mathbf{\nabla}_{\xb}^{(t)}$ and previous iterand $\xb^{(t)}$. These three components are converted into 6 channels, with two channels, per component, for the real and imaginary parts, respectively. This is propagated through the {GD-RegNet} as follows:
\begin{enumerate}
    \item First a 1-D convolution extracts the input features into a set of output channels\footnote{{The 1-D convolution empirically shown to provide improved results, compared to only using fully connected layers. Feature extraction from the OBMNet estimate and gradient enables a more robust GD regularization \eqref{eq_reg_one_bit_update_reg_gd_update}}.}
    \item The output of the 1-D convolution is flattened and passed through a fully connected network (FCN), consisting of three hidden layers. The output of the FCN is a vector in $\mathbb{R}^{2K}$, same as the $\hat{\xb}^{(t+1)}$
    \item A residual link from the OBMNet output $\hat{\xb}^{(t+1)}$ is added at the output of the {GD-RegNet}, generating the final iterand $\xb^{(t+1)}$.
    \item We normalize the final output $\xb^{(T)}$, analogous to \eqref{eq_obmnet_normalize}, as 
        \begin{equation}\label{eq_robnet_normalize}
            \xb^{(T)}\leftarrow \eta_{M}\,\frac{\xb^{(T)}}{||\xb^{(T)}||},
        \end{equation}
    where $\eta_{M}$ depends on the constellation order M\footnote{For lower order constellations, i.e., QPSK, we incorporate $\eta_{M}$ into the learning process during training, making it data-dependent. However, we have empirically observed that for higher order QAM, i.e., 16-QAM, this value should be fixed. On the whole, the difference between statically choosing $\eta_{M}$ and learning it from the data does not have any change in overall performance.}.
\end{enumerate}
The specific details of the parameters in each layer, for a general number of users K, are given in Table \ref{table_dnn_param}. 

The network training is carried out via minibatch gradient descent, with the chosen batch size $N_{\mathrm{train}}=32$. In order to train the {ROBNet} on {the set of randomly generated} Rayleigh channel matrices, each minibatch is generated from a different channel matrix $\Hb$, denoted by $\mathcal{B}_{\Hb}$. Based on the described system model \eqref{eq_unqunat_sig}-\eqref{eq_one_bit_quant}, the minibatch set is generated as $\mathcal{B}_{\Hb}=\{\bar{\xb}_{n},\bar{\zb}_{n},\bar{\yb}_{n}\}_{n=1}^{N_{\mathrm{train}}}$. We employ the modified loss function \eqref{eq_reg_one_bit_loss_func}, discussed in Sec. \ref{Sec_improved_loss}, to train the ROBNet. We practically implement minibatch gradient descent with the Adam update \cite{kingma2014adam} for each training minibatch to keep a check on the learning rate. {For regularization of DNN weights,} we utilize weight decay to further increase resilience by preventing exploding network weights.

%Implementing an iterative algorithm via an unfolded network, as we have demonstrated here, is one of the two approaches we have explored. We now investigate an alternate network strategy that can model the sequence of iterands $\{\xb^{(t)}\}_{t=0}^{T}$ as a time-series generated via the regularized GD algorithm. We thus turn towards recurrent neural networks to learn this time series pattern, resulting in a parametrically efficient network design.

\subsection{Recurrent one-bit DNN: OBiRIM}\label{Sec_obirim}
\begin{figure}[t]
	\centering
	\includegraphics[width=\linewidth]{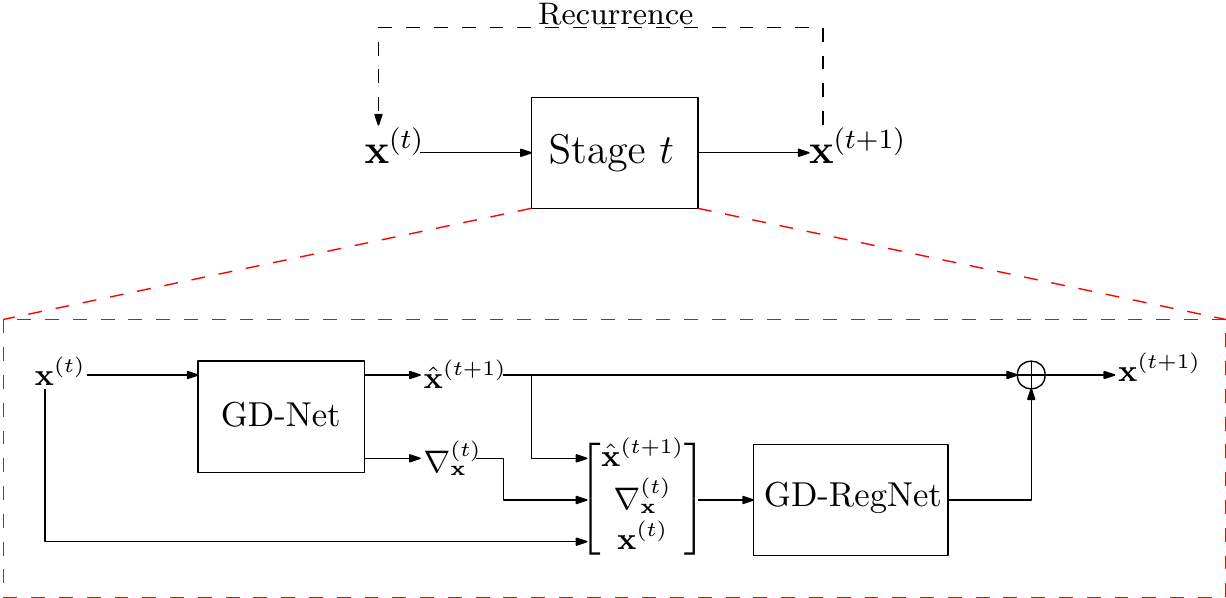}
	\caption{{Block diagram for the Projected-Regularized One-bit Recurrent Inference Machine (OBiRIM) Detector}}
	\label{fig_proj_obirim_overall}
\end{figure}
\noindent 
We now investigate an alternate network strategy that can model the sequence of iterands $\{\xb^{(t)}\}_{t=0}^{T}$ as a time-series generated via the regularized GD algorithm. We thus turn towards recurrent neural networks to learn this time series pattern, resulting in a parametrically efficient network design.

Recurrent neural networks (RNNs) have been one of the earliest DNNs to incorporate time-series information in pattern extraction for applications like speech and NLP \cite{chung2014empirical,young2018recent}. A specific class of these networks, the recurrent inference machines (RIMs), proposed in \cite{putzky2017recurrent}, have shown much success in medical imaging. The ability of the RIM architecture to parametrically model a prior distribution and as well as the optimization procedure is responsible for its superior performance over conventional approaches \cite{putzky2017recurrent,lonning2019recurrent}. The use of a parametric prior distribution as regularization, along with an iterative estimation algorithm of one-bit detection fits in perfectly with the strengths of the RIM framework. 

To this end, we implement our own version, the one-bit RIM (OBiRIM) for the regularized GD algorithm \eqref{eq_reg_one_bit_update}. The overall network structure for the OBiRIM is shown in Fig. \ref{fig_proj_obirim_overall}. Different from the ROBNet, the OBiRIM utilizes parameter sharing for the {GD-RegNet}, such that the same set of parameters $h_{\phi}$ are used for each GD iteration in \eqref{eq_reg_one_bit_update}. The presence of recurrent cells in the OBiRIM, stores the relevant estimation memory for the iterative GD algorithm and fine-tunes each OBMNet estimate $\hat{\xb}^{(t)}$ based on the system history. By sharing parameters among the different iterations and exploiting the system memory, this network is highly parameter efficient. To the best of our knowledge, the OBiRIM is the first recurrent DNN for detection of one-bit MIMO, which can be generalized for detection to any arbitrary Rayleigh fading channel. 

The overall regularized GD framework for the OBiRIM, as seen in Fig. \ref{fig_proj_obirim_overall}, is similar to the ROBNet, i.e., implementation of the regularized GD algorithm \eqref{eq_reg_one_bit_update}. We highlight some of the salient features of this network here below. 
\begin{itemize}
    \item The OBiRIM, implementing a $T$-step regularized GD algorithm, consists of $T$ temporal iterations. {At each Stage $t$ (see Fig. \ref{fig_proj_obirim_overall}), the data is sequentially processed through two phases}.
        \begin{enumerate}
            \item  {GD-Net - This is similar to GD-Net block in the ROBNet (see Fig. \ref{fig_proj_robnet_overall}).}
            \item {GD-RegNet - Different from the equivalent network of the ROBNet, this GD-RegNet incorporates DNN recurrence to temporally fine tune each estimate $\xb^{(t)}$. DNN memory enables temporal processing, while sharing parameters across different OBiRIM stages.}
        \end{enumerate}
    \item At each Stage $t$, a residual link from the {GD-Net output $\hat{\xb}^{(t)}$} is fed to the output of the {GD-RegNet}, thus imparting a stage-dependent correction to the unconstrained gradient step. 
\end{itemize}
We now provide the technical parameters {of this GD regularization}, along with the {general channel training}. 

\medskip
\noindent \textit{{GD-RegNet} structure and training:}
The {GD-Net output at each Stage $t$} is fed as 6 input channels to the {GD-RegNet}, similar to the ROBNet. Different from the series of {GD-RegNets} $\{h_{\phi}^{(t)}\}_{t=1}^{T}$ of the ROBNet, {the GD-RegNet} $h_{\phi}$ of the OBiRIM {is a single recurrent network} using gated recurrent units (GRU) to store the estimation memory. We choose the GRU as the recurrent block due to its ability to capture long and short term memory by resetting and updating the hidden state using the input sequence \cite{cho2014learning,chung2014empirical}.

The overall propagation of the input through the {GD-RegNet}, for each temporal Stage $t$, is given as:
\begin{enumerate}
    \item First, a 1-D convolution extracts the input features into a set of output channels. 
    \item The output of the convolution stage is  flattened and passed to the recurrent step of the {GD-RegNet}. This consists of two sequential GRU blocks, with the output hidden state of the first GRU cell passed as the input to the second GRU cell. The hidden states of both these recurrent cells are initialized to zero. 
    \item Post propagation through the two GRU cells, the output hidden state of the second GRU cell is flattened and passed to a FCN with four hidden layers, with the output of the FCN, having the same dimension as the OBMNet output $\hat{\xb}^{(t+1)}$.
    \item A residual link, from the OBMNet output $\hat{\xb}^{(t+1)}$, is added to the output of the FCN, similar to the ROBNet. 
    \item The normalization of the final estimate $\xb^{(T)}$ is carried out as in \eqref{eq_robnet_normalize}.
\end{enumerate}
The specific details of the {GD-RegNet} parameters of the OBiRIM, for a general number of users $K$, are given in Table \ref{table_dnn_param}.
The training data as well as the training parameters are the same as the that of the ROBNet (see Sec. \ref{Sec_robnet}). Further, the same improved loss function \eqref{eq_reg_one_bit_loss_func} is used to also train the OBiRIM network parameters. 

\begin{remark}\label{Remark_final_iter_loss}
    Both the ROBNet and OBiRIM are trained based on the loss function \eqref{eq_reg_one_bit_loss_func}, incorporating the final one-bit estimate $\xb^{(T)}$.
    Differently, the original RIM framework, introduced in \cite{putzky2017recurrent}, incorporates all the intermediate iterands $\xb^{(t)}$, with $t<T$, in the evaluation of the MSE loss.
    Although the analysis stemming from the explicit incorporation of these intermediate iterands in the final loss function falls outside the scope of this work, we have utilized this strategy for a different context of one-bit detection. This analysis for one-bit MIMO is left for our future work.
\end{remark}

\section{Experimental Results}\label{Sec_exp_results}
\noindent We now evaluate our regularized networks ROBNet and OBiRIM. First we describe the simulation setup, followed by the  results of the various tests along with comments.

\smallskip
\noindent \textit{Simulation setup}

\noindent We evaluate the detector on two different M-QAM constellations with different channels, user, BS antennas and input $\mathrm{SNR}=\frac{\mathbb{E}(||\Hb\xb||^{2})}{\mathbb{E}(||\nb||^{2})}$ parameters:
\begin{enumerate}[label=(\roman*)]
    \item The QPSK constellation with $K=4$ users, $N=32$ BS antennas and $\mathrm{SNR}$ in the range $-5$ to $35$ $\mathrm{dB}$.
    \item The 16-QAM constellation with $K=8$ users, $N=128$ BS antennas with $\mathrm{SNR}$ in the range $10$ to $45$ $\mathrm{dB}$.
\end{enumerate}
{Both the simulations setups (i) and (ii) follow the standard simulations conducted in \cite{choi2016near,nguyen2021linear,Ho2022Thesis}.} For both the constellation cases, a Rayleigh fading channel $\Hb$ is considered with each entry chosen from the $\mathcal{CN}(0,1)$ distribution. Unless otherwise stated, we assume perfect channel state information (CSI) available at the BS.

\smallskip
\noindent \textit{Performance benchmarks}

\noindent We benchmark our algorithms against the existing model-based and DNN-aided one-bit detectors for state-of-the-art detection. For the simulation setup \textit{(i)}, as described in the paragraph above, we lower bound the BER by the maximum-likelihood detector (ML Detector). Using the exhaustive constellation search, this method grows exponentially with each added user as well as increase in modulation order. However, this presents the best recovery possible, directly solving the constrained optimization problem \eqref{eq_one_bit_ml_sigmoid}. ML detection for 16-QAM (simulation setup \textit{(ii)}) presents as a much larger computational complexity for our scale of the simulation setup considered, and is hence not evaluated. The OBMNet \cite{nguyen2021linear} is used as the main benchmark, on which we propose improving, by means of the regularized GD \eqref{eq_reg_one_bit_update}. We also provide the performance of the n-ML algorithm, from \cite{choi2016near}, to benchmark against the GD-based detector using cdf-based likelihood \eqref{eq_one_bit_ml_prob}. For testing the general channel detection performance we also benchmark our algorithm against the FBM-DetNet \cite{nguyen2022deep}, implemented for the same number of iterations as the OBMNet.

\begin{remark}\label{remark_dnn_benchmark}
    The work in \cite{Ho2022Thesis} extensively tests end-to-end learning via different DNNs like Resnets, Densenets and Hypernetworks for one-bit detection. However, the presented results in this work show the robust model-based OBMNet to exceed the performance of these networks.
    Thus, we have omitted the inclusion of these end-to-end learning approaches for benchmarking our regularized one-bit detection approach.
\end{remark}

%\medskip
\noindent \textit{Network and model parameters}

\noindent {Consistent with the benchmarks established in \cite{nguyen2021linear}}, the OBMNet is run for ten and fifteen iterations $(T\in\{10,15\})$ for simulation setups \textit{(i)} and \textit{(ii)}, respectively. The n-ML method is executed for {a maximum of} $T=500$ iterations, with a step size of $0.001$, {to ensure convergence}.
The network parameters and training details for our proposed networks - ROBNet and OBiRIM, have been provided in Sec. \ref{Sec_robnet} and \ref{Sec_obirim}. In contrast to the higher number of iterations for the given benchmarks above, we execute both the ROBNet and the OBiRIM for only five and ten GD iterations $(T\in\{5,10\})$ for simulation setups \textit{(i)} and \textit{(ii)}, respectively. Thus the added utility of the regularized GD algorithm also presents as an advantage in reducing the number of GD iterations.
To avoid overloading the networks for large SNR ranges during training, the proposed networks are trained for a single intermediate SNR ($15$ $\mathrm{dB}$ for simulation \textit{(i)} and $25$ $\mathrm{dB}$ for simulation \textit{(ii)}) and tested on the entire range mentioned above. A similar strategy for training unfolded and recurrent neural networks was used in \cite{sant2022deep}.

\subsection{{Intrinsic testing of DNN-aided regularized GD}}\label{Sec_intrinsic_testing}
\begin{figure}[t]
	\centering
	\includegraphics[width=0.8\linewidth]{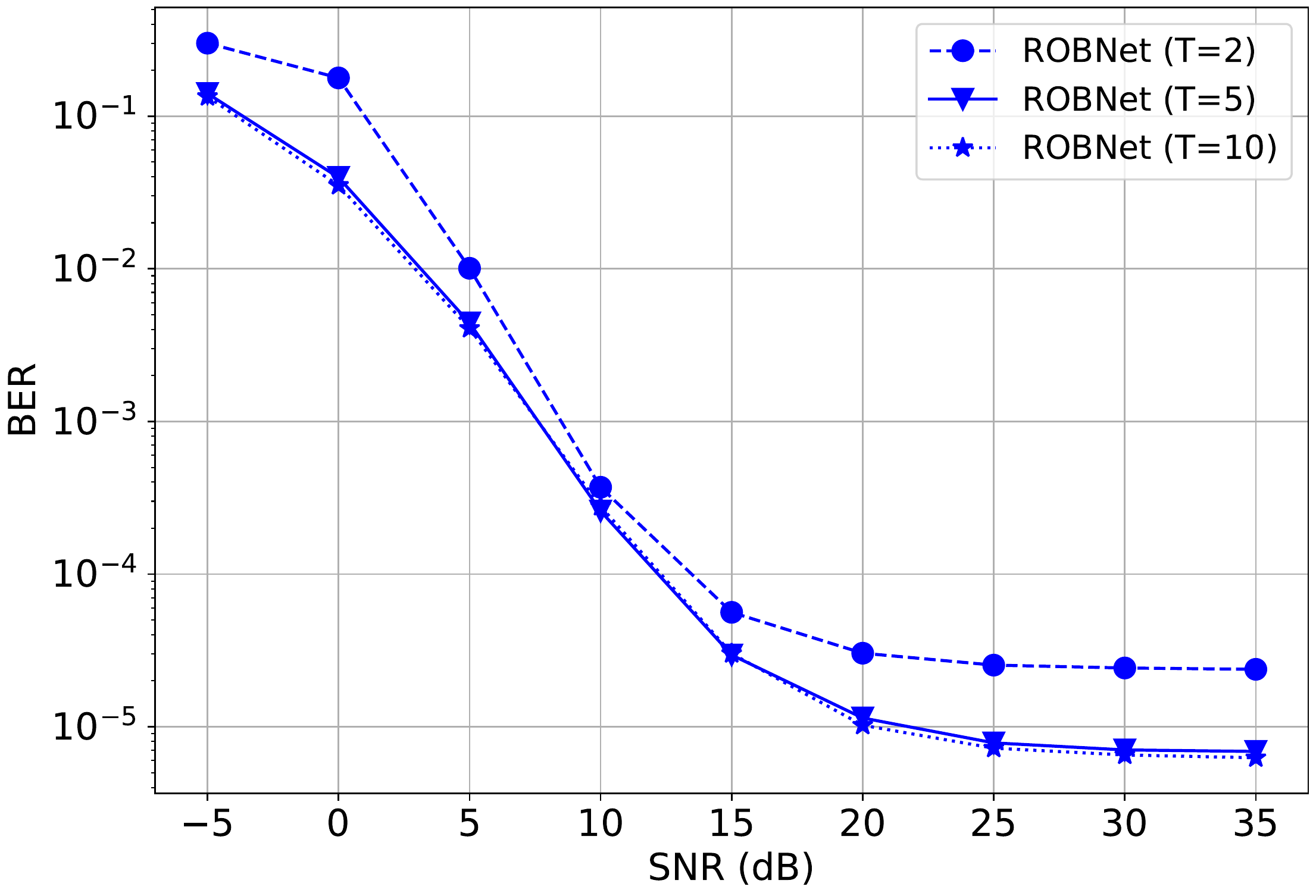}
	\caption{Testing the ROBNet for different number of stages $T$ for QPSK transmitted symbols, with $N=32$, $K=4$}
	\label{fig_num_layers_test_qpsk}
\end{figure}
\begin{figure}[t]
	\centering
	\includegraphics[width=0.8\linewidth]{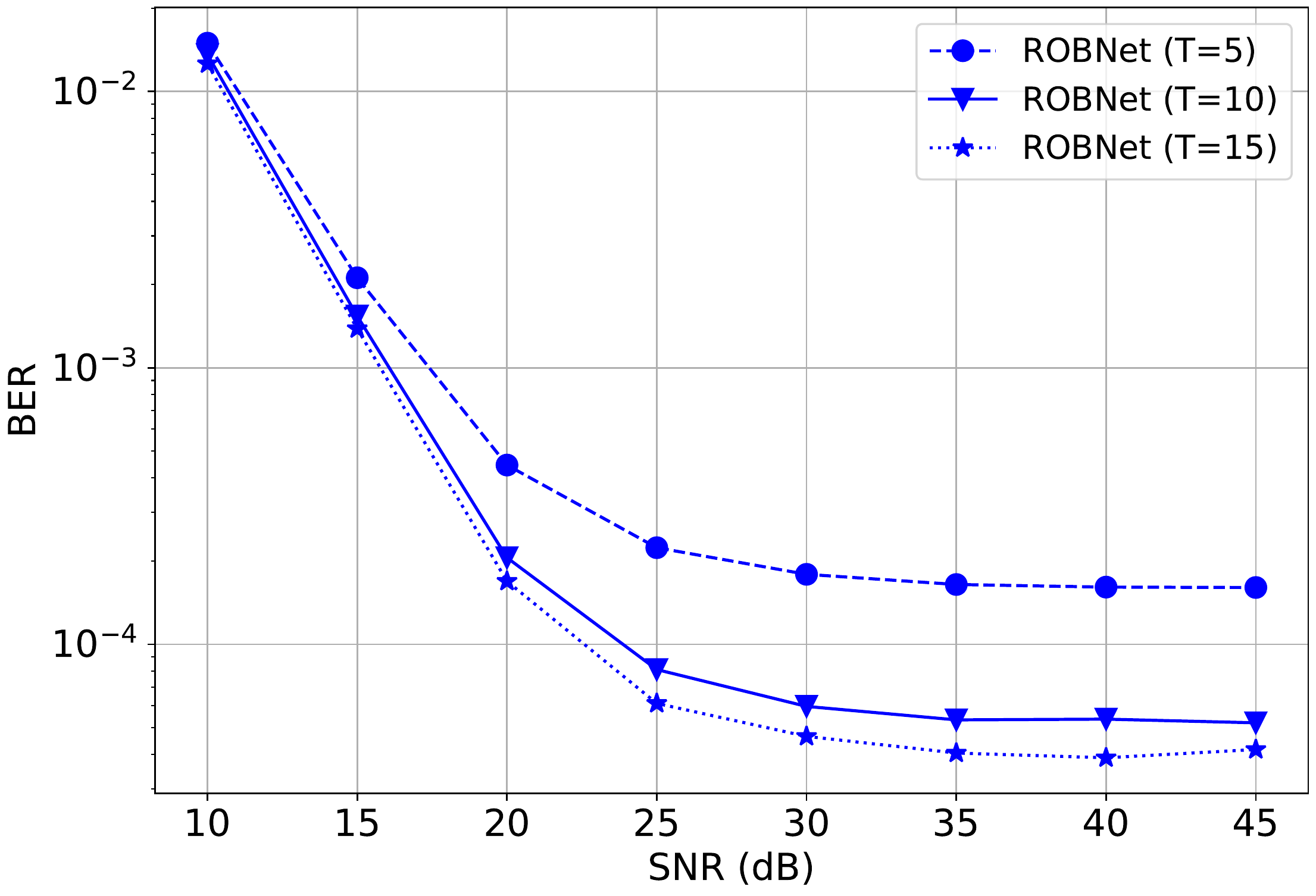}
	\caption{Testing the ROBNet for different number of stages $T$ for QPSK transmitted symbols, with $N=128$, $K=8$}
	\label{fig_num_layers_test_qam16}
\end{figure}
\noindent {We begin by testing the performance of the networks implementing the regularized GD update \eqref{eq_reg_one_bit_update}. In particular, we test the performance by varying the network parameters, i.e., the number of network stages $T$. For the considered test, we evaluate the performance of the unfolded network, the ROBNet. 
\subsubsection{Effect of number of stages $T$}
We train and test the ROBNet for different stages $T$, corresponding to number of regularized GD iterations. The plots given in Fig. \ref{fig_num_layers_test_qpsk} portray the BER performance for the QPSK transmitted symbols. Based on the performance plots, There is a marked gain in performance from $T=2$ to $5$ stages, with saturation in performance beyond this. By increasing the number of layers beyond a certain limit, we increase network complexity and runtime with extremely marginal gains in performance. This is also supported by the results for the 16-QAM constellation, as shown by the plots in Fig. \ref{fig_num_layers_test_qam16}. Here too, a significant performance boost is observed as we increase from $T=5$ to $10$, with subsequent increase in the number of stages only marginally increasing performance. Based on the observed recovery results, we utlize $T=5$ and $T=10$ layers for QPSK and 16-QAM symbols, respectively.
}

\subsection{Recovered constellation}\label{Sec_recovered_const}
\begin{figure}[t]
	\centering
	\includegraphics[width=\linewidth]{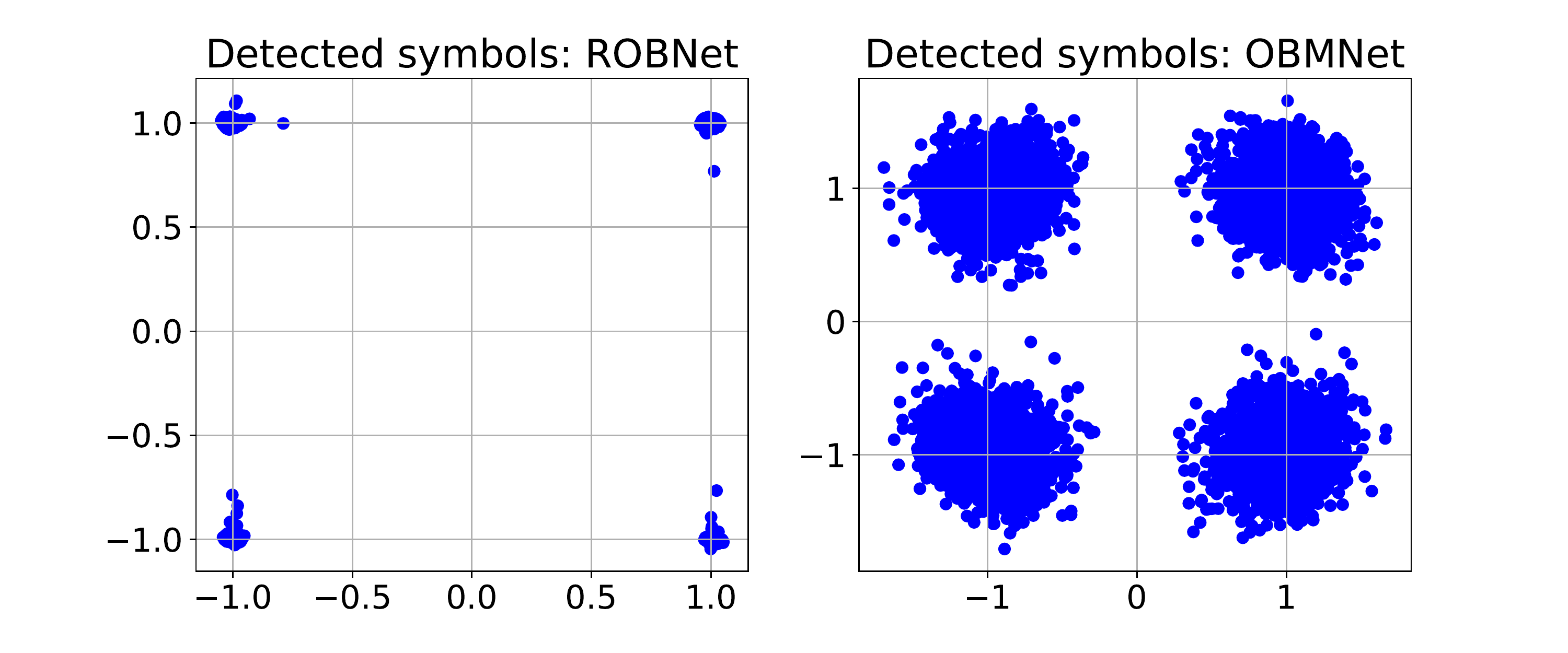}
	\caption{Recovered QPSK constellation for ROBNet compared to OBMNet \cite{nguyen2021linear}, with $N=32$, $K=4$ (red dots represent incorrectly detected symbols)}
	\label{fig_const_recovered_qpsk}
\end{figure}
\begin{figure}[t]
	\centering
	\includegraphics[width=\linewidth]{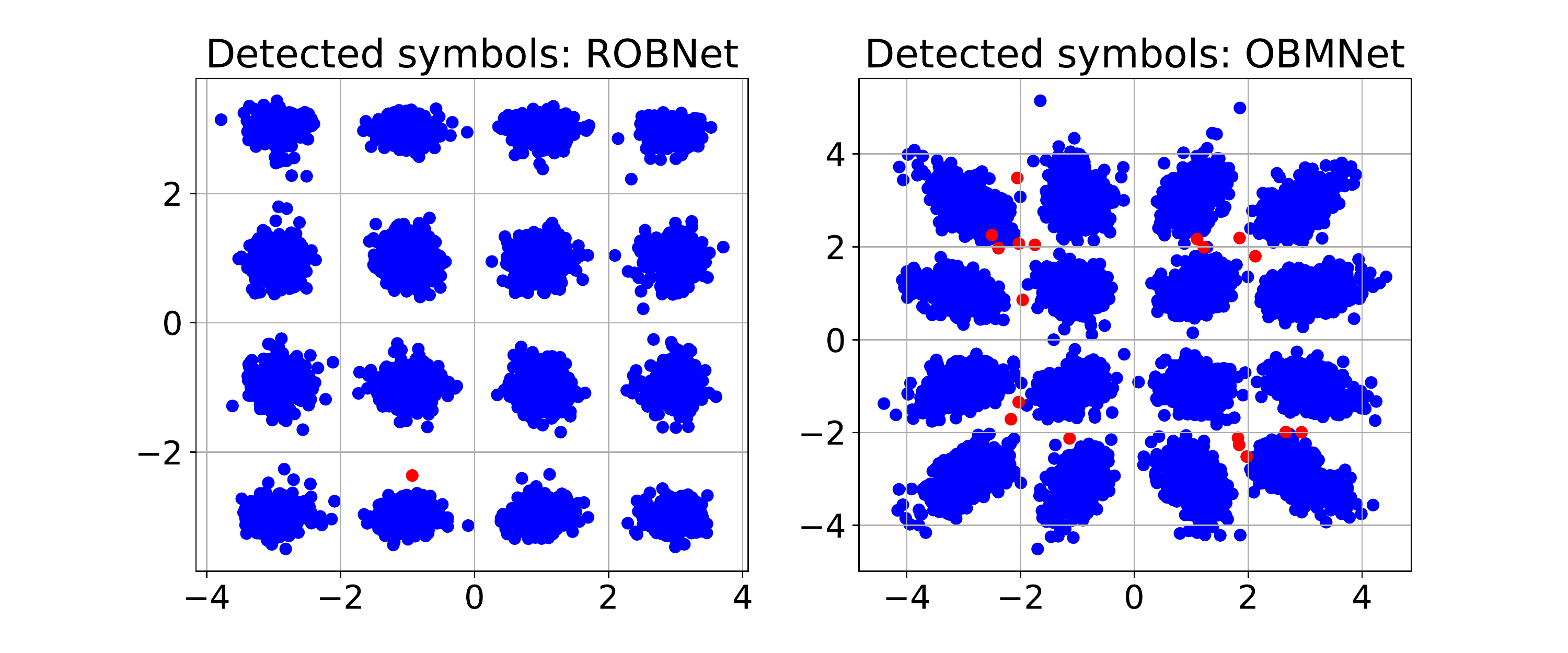}
	\caption{Recovered 16-QAM constellation for ROBNet compared to OBMNet \cite{nguyen2021linear}, with $N=128$, $K=8$ (red dots represent incorrectly detected symbols)}
	\label{fig_const_recovered_16qam}
\end{figure}
\noindent We qualitatively analyze the recovered M-QAM constellation symbols, comparing these to the recovered symbols from the OBMNet, which doesn't utilize any additional regularization. The symbol recovery is demonstrated for the training SNR of the networks, i.e., $15\,\mathrm{dB}$ for QPSK and $25\,\mathrm{dB}$ for 16-QAM. The recovered QPSK symbols are given in Fig. \ref{fig_const_recovered_qpsk}. As can be seen from these plots, the OBMNet results in recovered symbols with a larger cluster spread. The combination of the increased network expressivity for the ROBNet, along with the constellation-aware network loss function \eqref{eq_reg_one_bit_loss_func}, results in much sharper recovered symbol clusters. 

The symbol recovery for 16-QAM constellation presents the more stark contrast on the effect of the regularized GD method \eqref{eq_reg_one_bit_update}. Although the OBMNet is able to effectively recover the 16-QAM symbols from the one bit data, the cluster shapes are non-homogeneuous in the symbol power. As can be seen from the density of incorrectly detected symbols (red scatter points), this non-homogeneity results in more incorrectly detected symbols. The regularization introduced by the ROBNet, in contrast, presents a more homogeneous recovered constellation, irrespective of the 16-QAM symbol powers. As visually evident, this is responsible for fewer incorrectly detected symbols.

Following the qualitative visual analysis of the recovered symbols, we now move on to the quantitative analysis.

\subsection{Detection for single Rayleigh-fading channel}\label{Sec_exp_single_chan}
\begin{figure}[t]
	\centering
	\includegraphics[width=\linewidth]{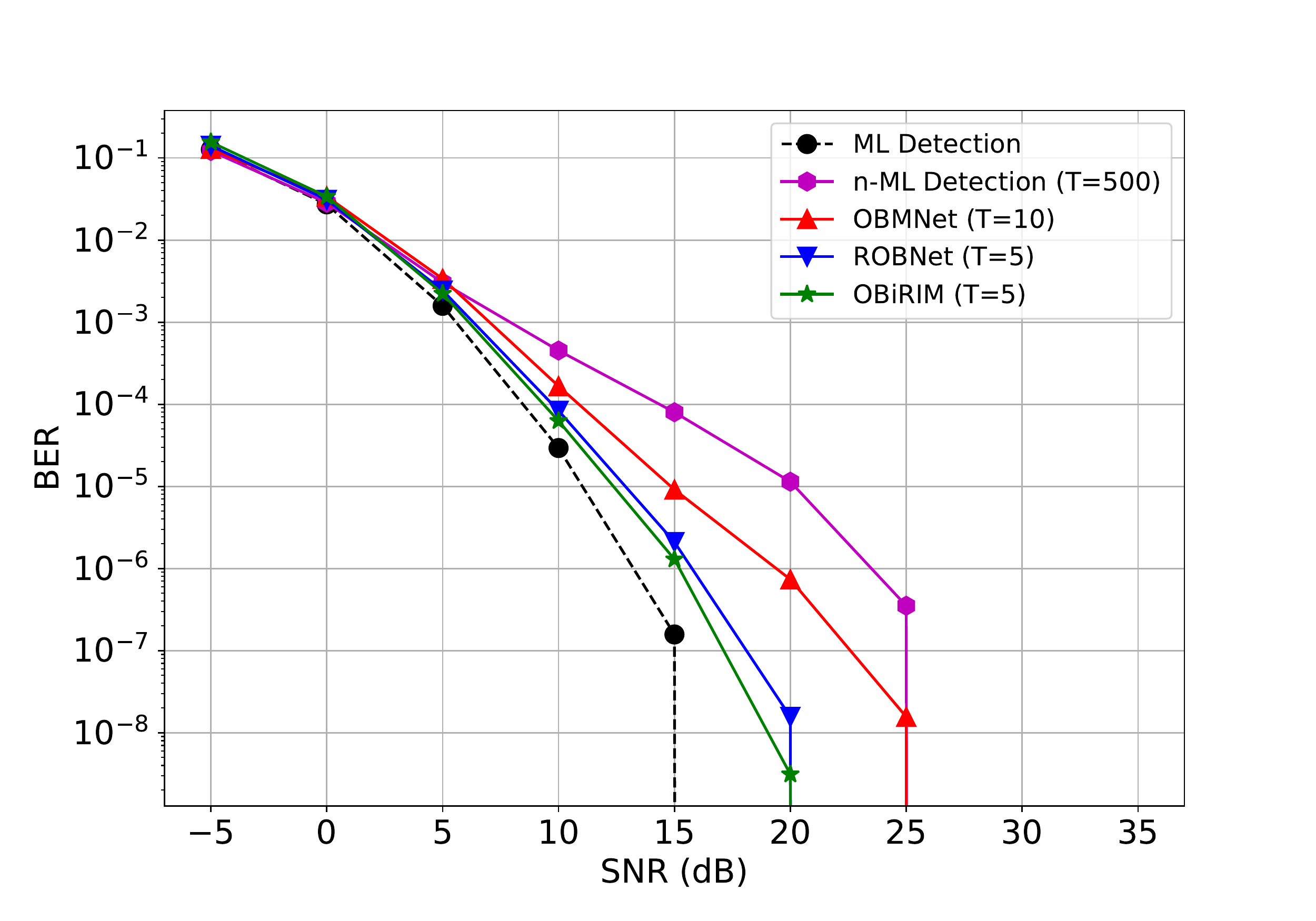}
	\caption{Performance comparison of improved networks for channel-specific detection for QPSK constellation with number of antennas $N=32$ and the number of users $K=4$.}
	\label{fig_results_qpsk_H_mat_05}
\end{figure}
\begin{figure}[t]
	\centering
	\includegraphics[width=\linewidth]{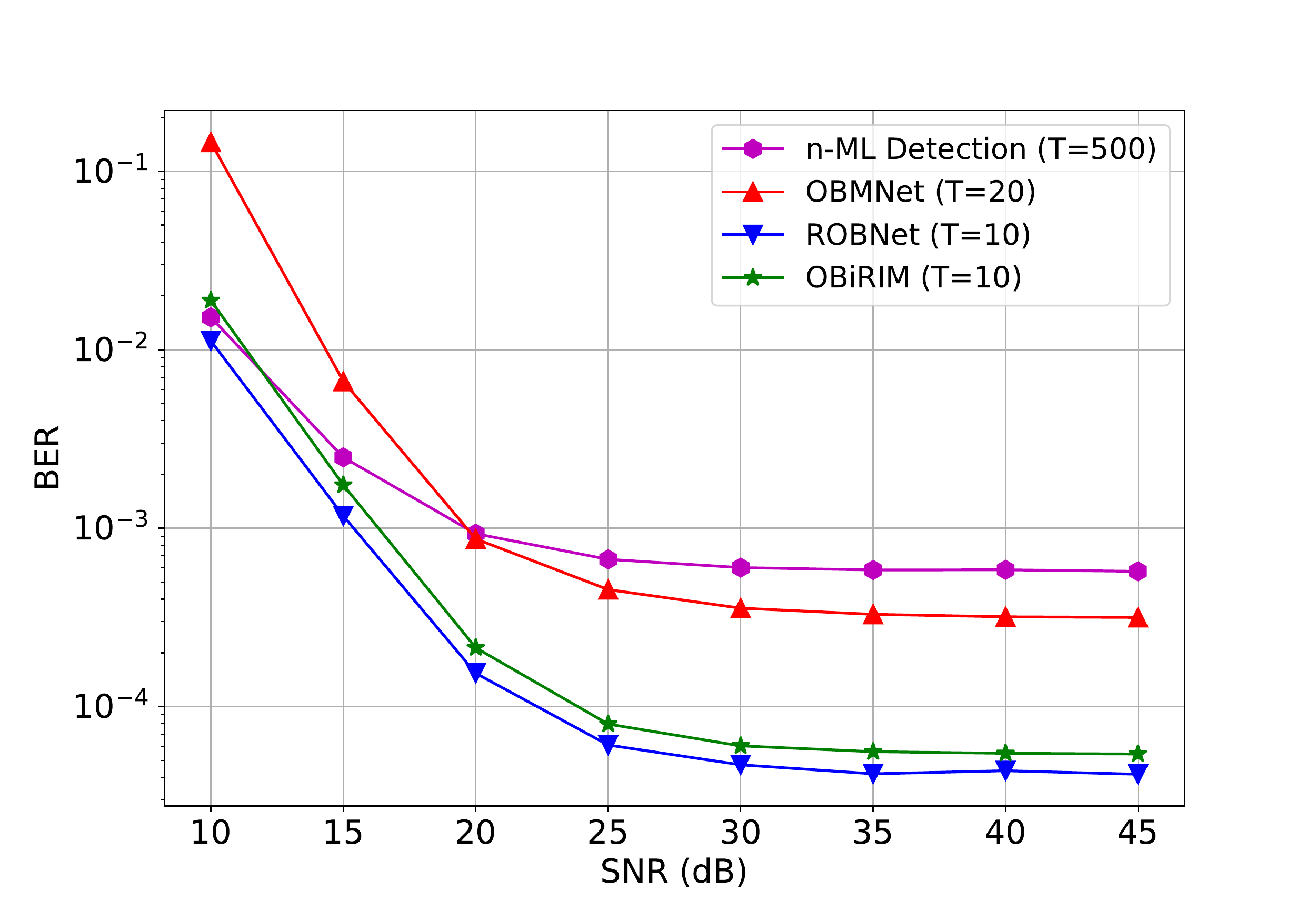}
	\caption{Performance comparison of improved networks for channel-specific detection for 16-QAM constellation with number of antennas $N=128$ and the number of users $K=8$.}
	\label{fig_results_16qam_H_mat_05}
\end{figure}
\noindent Through this test we demonstrate the strength of our proposed approach, when implemented for channel-specific detection. As stated in Sec. \ref{Sec_Intro}, most conventional end-to-end DNN-based detectors, both for unquantized as well as one-bit received data \cite{pratik2020re,khani2020adaptive,he2020model,khobahi2021lord,kim2020machine,balevi2019one}, are trained and tested for a single channel. 
{Such detectors are applicable to highly static and directional channels, with minimum CSI variation. Real-world channels, like Rayleigh-fading channels, are more dynamic; robust detector design should thus be channel state-invariant, trained on the entire set of random Rayleigh-fading channels and avoiding the need to be retrained for each new CSI matrix.}
Prior to testing the proposed networks in this work on the entire distribution of Rayleigh-fading channels, we perform the channel-specific detection to ascertain the performance for this widely utilized model by different works. In the context of DNN design, this test is akin to the overfitting test.
The different networks and approaches are trained and tested on a single channel $\Hb$, sampled from the distribution of Rayleigh-fading channels. Further, we normalize the columns of the channel matrix $\Hb$ and scale it by the number of antennas $N$, ensuring each user receives the same channel power. 

The channel-specific BER performance for QPSK symbols is shown in Fig. \ref{fig_results_qpsk_H_mat_05}. As seen from this plot, all the networks and the algorithms approach very low BER values when trained and tested with a very well conditioned channel with equal power distribution among the corresponding users \footnote{Equal per-user channel power is especially important for joint GD-based detection. We analytically validate this, in detail, through our future work.}. However, such ideal performance requires overfitting the networks for a given channel model, which presents extensive practical challenges. As seen from the plots, the OBMNet, with its improved sigmoid likelihood formulation exceeds the n-ML approach, further highlighting the utility of this likelihood formulation. As can be seen from the results, both our proposed networks exceed the OBMNet performance for the channel-specific detection, approaching the ideal ML-detection.
This further enforces the utility of the proposed regularized GD algorithm and the constellation-aware loss function. 

The channel-specific BER performance for the 16-QAM symbols is shown in Fig. \ref{fig_results_16qam_H_mat_05}. Here too, all the networks and algorithms are trained and tested on a single channel matrix, with equal per-user channel power. Based on the results in this plot, the contrast in performance between the regularized one-bit GD and the competing algorithms is more starkly visible, highlighting the strength of this strategy for higher order M-QAM constellations. The presence of a non-zero BER floor for all algorithms (as compared to Fig. \ref{fig_results_qpsk_H_mat_05}), stems from the more challenging case of recovering higher order constellation symbols from one-bit data. 

%However, the approach of training networks on a per-channel basis is useful for very static channel scenarios. Training the networks anew, on observing a different channel matrix markedly increases the access phase for the detector. as well as pilot utilization. As highlighted in Sec. \ref{Sec_gen_detect}, we design and train the neural detector on the Rayleigh channel ensemble, avoiding the need to retrain the network on observing a new channel response, with the performance analyzed in the subsequent sub-sections.

\subsection{Detection for general channel}\label{Sec_exp_gen_chan_results}
\begin{figure}[t]
	\centering
	\includegraphics[width=\linewidth]{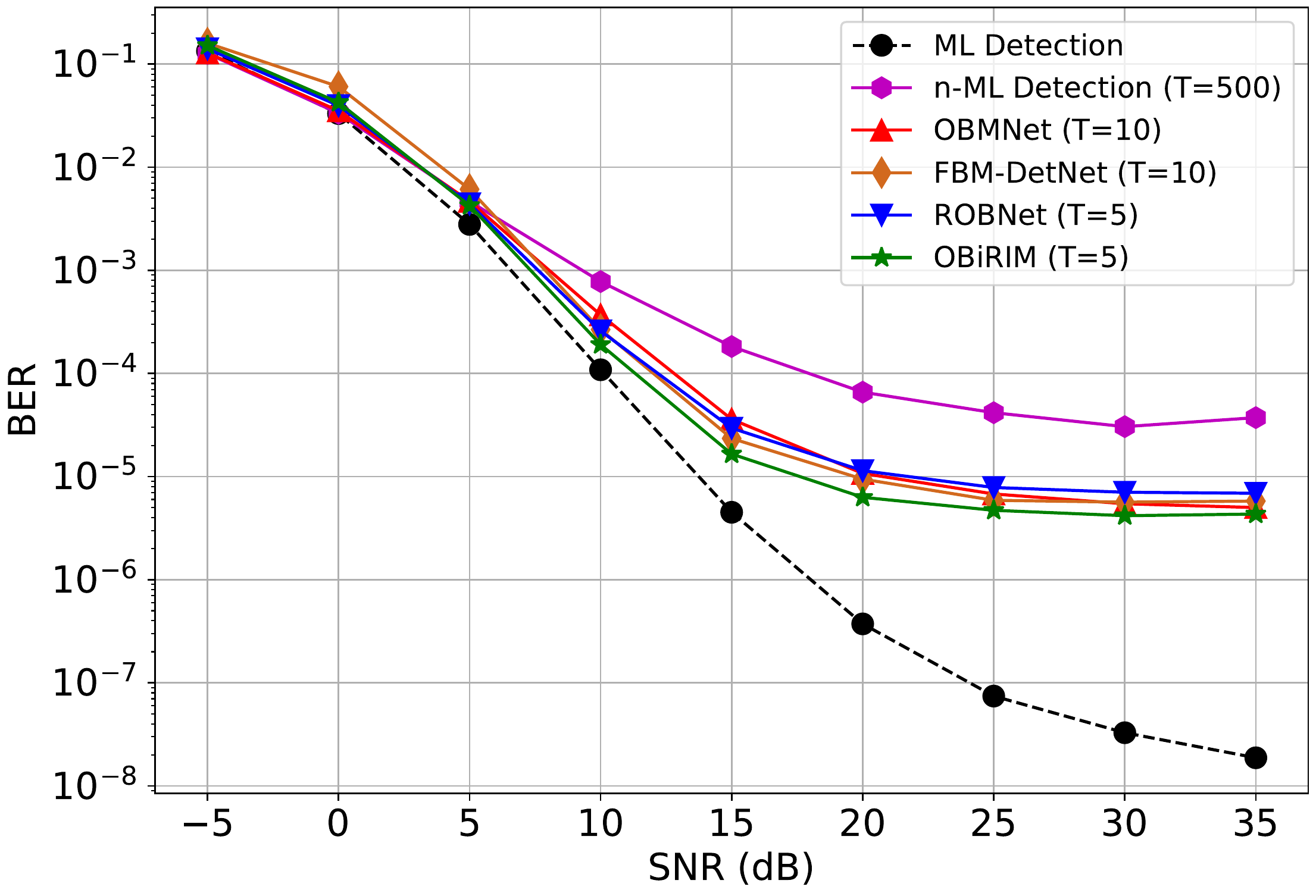}
	\caption{Performance comparison of improved networks for general channel detection for QPSK constellation with number of antennas $N=32$ and the number of users $K=4$.}
	\label{fig_results_qpsk_gen_channel}
\end{figure}
\begin{figure}[t]
	\centering
	\includegraphics[width=\linewidth]{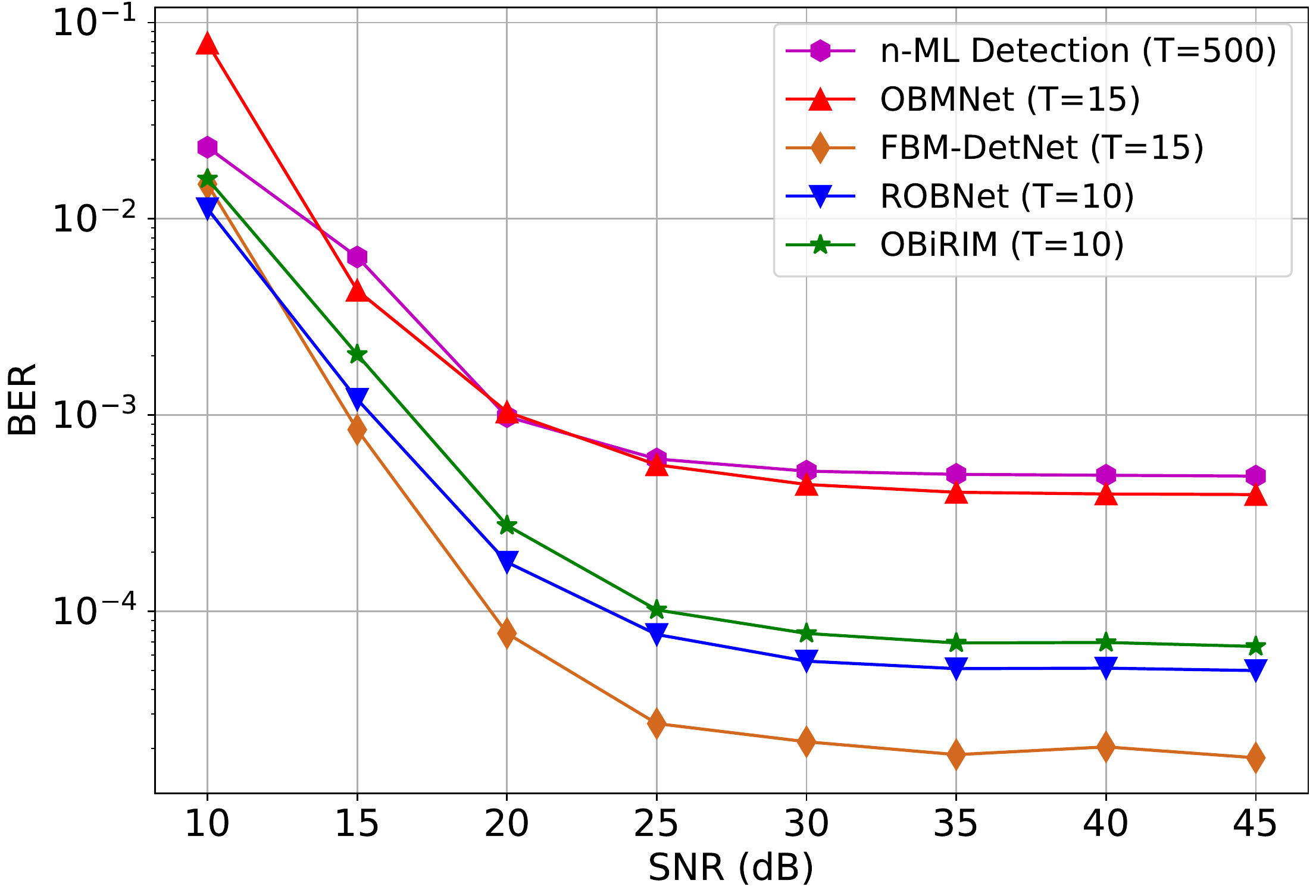}
	\caption{Performance comparison of improved networks for general channel detection for 16-QAM constellation with number of antennas $N=128$ and the number of users $K=8$.}
	\label{fig_results_qam16_gen_channel}
\end{figure}
\noindent We now present the results for the networks trained and tested on the {set of all} Rayleigh-fading channel matrices $\Hb$, {by randomly sampling from this distribution for each training minibatch}. Once the networks have been trained in this manner, they do not need to be fine-tuned or re-trained for each new channel matrix, thus acting as general Rayleigh channel detectors.
%The benchmark DNN-based detector {[add ref]} is trained and fine-tuned for a particular channel matrix and hence cannot be used for general channel detection. We thus benchmark our proposed detectors against the model based algorithms: n-ML detection and OBMNet.

The general channel BER performance for the QPSK symbols is given in Fig. \ref{fig_results_qpsk_gen_channel}. As seen from the performance plots, our proposed DNN-based detectors, ROBNet and OBiRIM, are able to generate the same performance as the OBMNet and the FBM-DetNet for the QPSK symbols with much fewer GD iterations. In addition, as seen from the recovered constellations plots in Sec. \ref{Sec_recovered_const}, these networks generate sharper constellation clusters with much smaller cluster spread. Although this does not directly translate to improved BER performance for lower order constellations like QPSK, it does bode advantageous for higher order constellations.

The BER performance for the 16-QAM constellation symbols is shown in Fig. \ref{fig_results_qam16_gen_channel}. As can be seen from these plots, the improved regularization framework directly translates to an improved relative BER performance, as compared to the OBMNet, for the higher order 16-QAM constellation symbols. Further, the BER performance, especially the high-SNR BER floor, for general channel detection is similar to that of the channel-specific detection, seen in Fig. \ref{fig_results_16qam_H_mat_05}. This can be attributed to the channel hardening effect seen by increasing the number of receiver antennas, improving the overall channel conditioning for any general 16-QAM channel. However, we observe that the quantization-specific learnable projection of the FBM-DetNet outperforms both the ROBNet and the OBiRIM. The sharper learnable quantization to the M-QAM symbols is responsible for lower cluster spreads. The ROBNet and OBiRIM are unable to sharpen constellation clusters beyond a certain limit and the hence under-perform in BER to the FBM-DetNet.

%{One possible way confirming channel hardening effect and also the ability of our approach to learn the general channel behavior is to conduct a test in which a network is trained on a single well-conditioned channel and tested on any general channel.}

\subsection{Detection for general channel - Noisy channel estimate}
\begin{figure}[t]
	\centering
	\includegraphics[width=\linewidth]{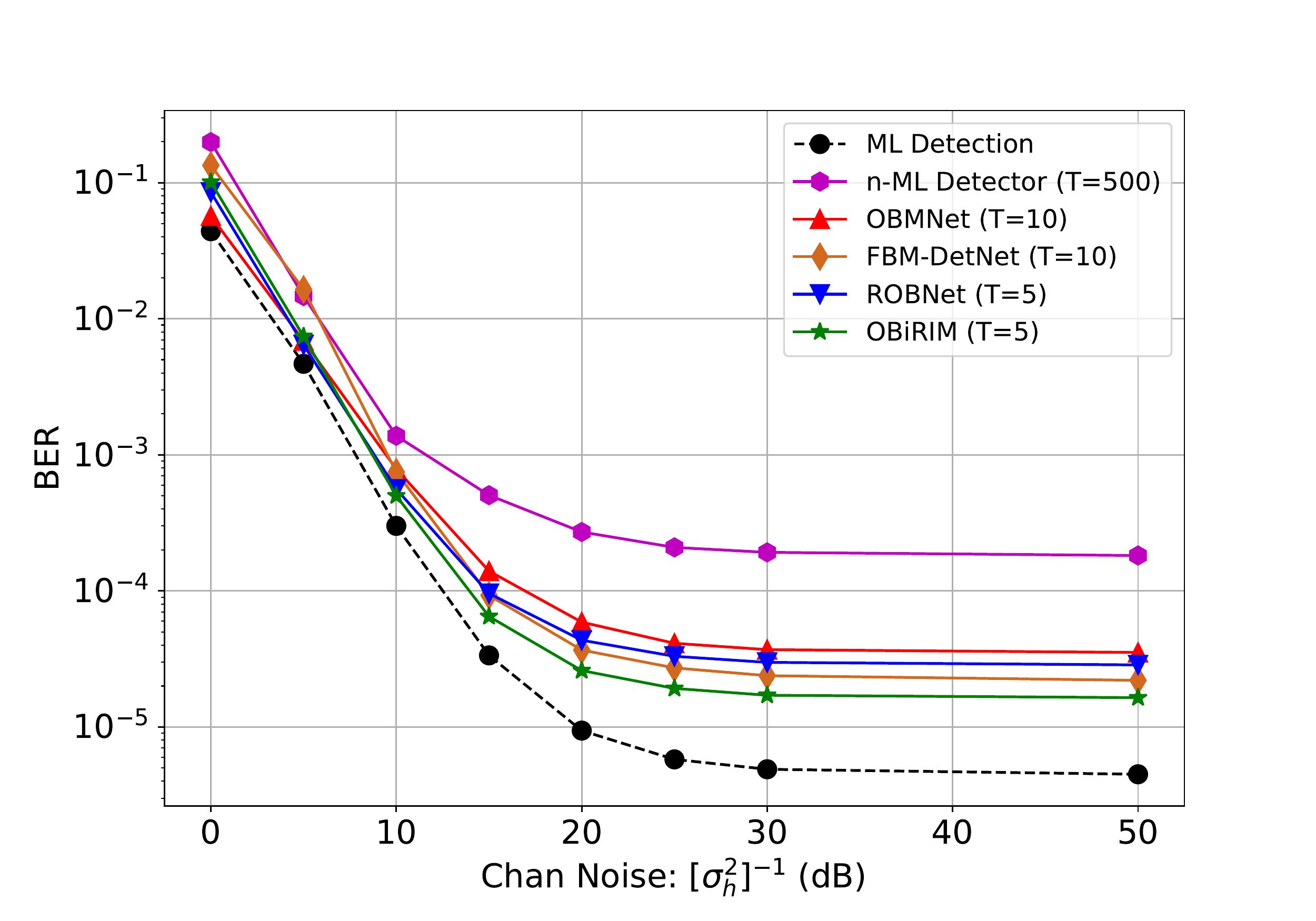}
	\caption{Performance comparison of improved networks for general channel detection with imperfect CSI for QPSK constellation with number of antennas $N=32$ and the number of users $K=4$.}
	\label{fig_results_qpsk_gen_channel_imperfect_csi}
\end{figure}
\begin{figure}[t]
	\centering
	\includegraphics[width=\linewidth]{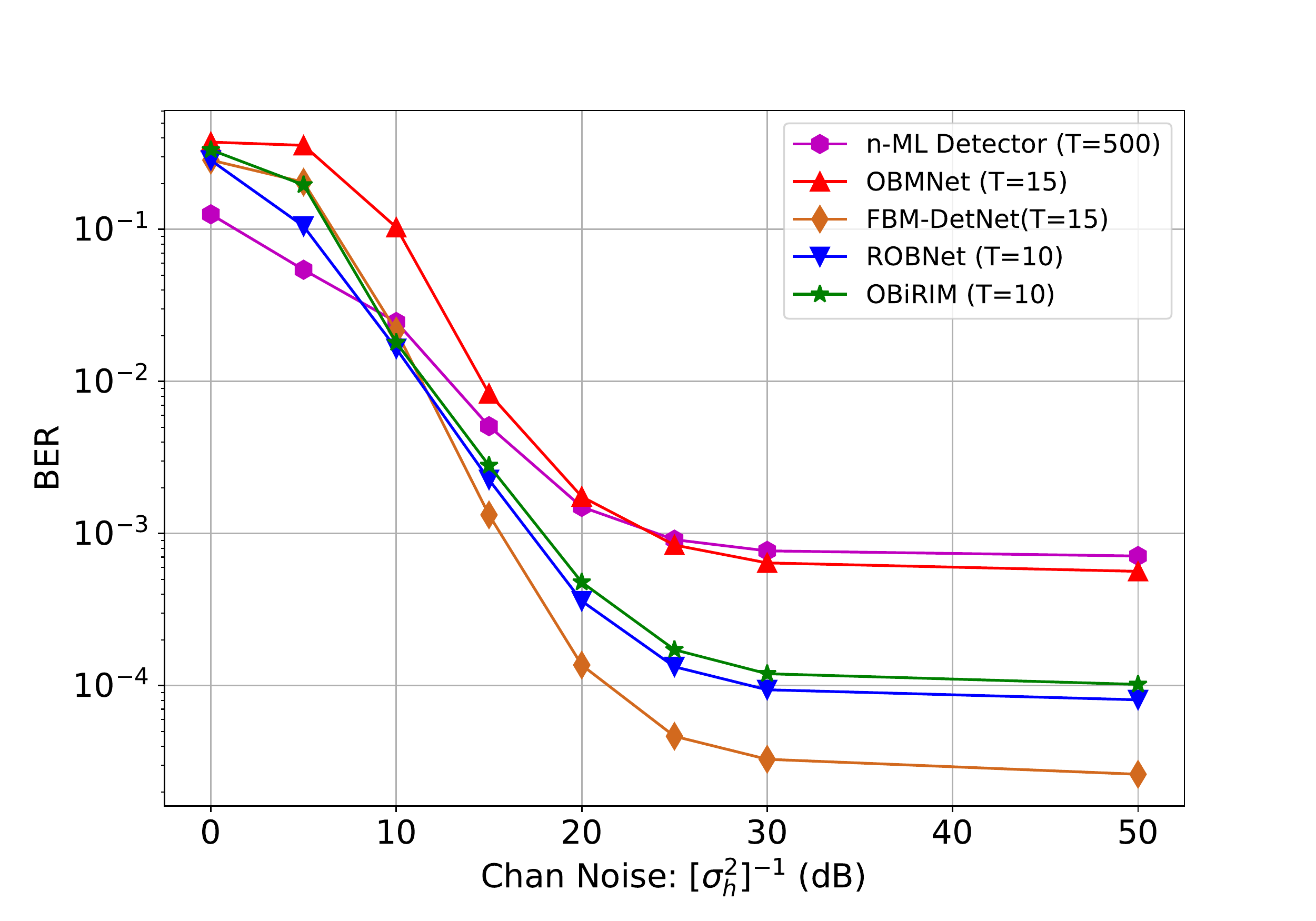}
	\caption{Performance comparison of improved networks for general channel detection with imperfect CSI for 16-QAM constellation with number of antennas $N=128$ and the number of users $K=8$.}
	\label{fig_results_16qam_gen_channel_imperfect_csi}
\end{figure}
\noindent The different model-based and DNN-based approaches described above rely on perfect channel estimates. However, practical systems introduce an estimation error in acquiring the channel state information. Although different channel estimation algorithms have been studied for one-bit systems \cite{stockle2016channel,liu2017one,sant2020doa}, we model the channel estimation via a general estimation error. The estimated Rayleigh fading channel is modeled as $\hat{\Hb} = \Hb + \Delta\hb$. Here the introduced estimation error is modeled as an additive complex Gaussian with each term $\begin{bmatrix}\Delta\hb\end{bmatrix}_{i,j}$ drawn from the distribution $\mathcal{CN}(0,\sigma_{\hb}^{2})$. We analyze the BER performance on the pre-trained networks with perfect CSI as a function of the introduced channel noise $\sigma_{\hb}^{2}$. All the trainable networks in the subsequent performance comparison are trained on perfect CSI, {setting a uniform reference point for all networks}, but tested on noisy CSI. {Through this test, we assess the inherent network resiliency for both the ROBNet as well as OBiRIM, compared to other benchmarks.}

The BER performance for the QPSK constellation symbols, as a function of this estimation noise $\sigma_{\hb}^{2}$ is given in Fig. \ref{fig_results_qpsk_gen_channel_imperfect_csi}. The detection performance is analyzed at the training SNR for QPSK symbols, i.e., $15\,\mathrm{dB}$.
As can be seen from these plots, both the regularized GD networks, i.e., the ROBNet and the OBiRIM, are more resilient to channel estimation noise, as compared to the unregularized OBMNet. Further, the FBM-DetNet also performs comparably to the ROBNet and OBiRIM. The small performance gap among all these algorithms goes on to further highlight the strength of the original OBMNet framework for lower order M-QAM constellations. Consistent with the results of Fig. \ref{fig_results_qpsk_gen_channel}, we observe marginal improvement over the OBMNet framework with additional regularization, for lower order M-QAM constellations.
However, increasing constellation order brings out the increased resilience of our proposed approach over the OBMNet. 

The BER performance for 16-QAM constellation symbols as a function of the added channel estimation error is provided in Fig. \ref{fig_results_16qam_gen_channel_imperfect_csi}. The detection performance is analyzed at the training SNR for 16-QAM symbols, i.e., $25\,\mathrm{dB}$.
There is a markedly increased performance gap in the performance of the regularized GD approach (both ROBNet and OBiRIM) for the 16-QAM constellation symbols.
The increased network expressivity and training of our proposed approach enables accommodation of CSI estimation errors, in spite of the presence of higher order constellation symbols. However, the FBM-DetNet outperforms both the ROBNet as well as the OBiRIM in the resilience to channel estimation noise. As stated in Sec. \ref{Sec_exp_gen_chan_results}, this is attributed to reduced cluster spread generated by the quantization-based projection to M-QAM symbols of the FBM-DetNet.

We thus infer that the combination of the general parametric regularization, improved loss function and {training on multiple Rayleigh-fading channel matrices}, makes these extremely robust, over the unregularized OBMNet, one-bit detection networks. The observed resilience to channel estimation errors enables the use of these networks in conjunction with standard one-bit channel estimation algorithms, without affecting detection performance. {Additionally, the proposed networks do not need to be separately trained for noisy channel estimates; existing ideally trained networks can be directly used with noisy channel data.}

\begin{remark}\label{remark_comparing_robnet_obirim}
    {Based on the observed results from Figs. \ref{fig_results_qpsk_H_mat_05}-\ref{fig_results_16qam_gen_channel_imperfect_csi}, we can observe a difference in behavior for the unfolded ROBNet and recurrent OBiRIM. In particular, the OBiRIM is shown to perform marginally better for the lower order QPSK, whereas the ROBNet performs marginally better for the higher order 16-QAM. This highlights an important trade-off between \textit{(i)} Capturing correlation through system memory and, \textit{(ii)} Network expressivity through number of parameters. For the simpler QPSK system model, the system memory, through DNN recurrence, is slightly more efficient at capturing correlation among the different intermediate iterates $\{\xb^{(t)}\}_{t=1}^{T}$. This is translates to marginally better performance for the simpler QPSK case. However, as we increase the constellation order, the recovery requires more network expressivity. Increasing the number of iterations the ROBNet increases the number of sub-networks and thus the number of trainable parameters. On the other hand, the OBiRIM, with more number of iterations, retains the same number of parameters due to parameter sharing. For the higher order 16-QAM symbols, we observe that the network expressivity and number of parameters wins out over the ability of the OBiRIM to capture correlation (with the same number of parameters). Thus the ROBNet now marginally outperforms the OBiRIM\footnote{{Detailed analysis of this trade-off between recurrence and parameter richness falls outside the scope of this work.}}.}
\end{remark}

\section{Conclusions}\label{Sec_conclusions}
\noindent In this work we have proposed a regularized one-bit neural detector based on a novel regularized GD-based strategy for improving on the state-of-the-art OBMNet. The learnable DNN-based regularization is effectively able to improve on the OBMNet estimate on a per-iteration basis. To this end, we have developed two unique regularization networks: \textit{(i)} ROBNet, using an unfolded DNN architecture, and \textit{(ii)} OBiRIM, using a RIM-based architecture. We also developed a novel constellation-aware loss function for DNN training, through which we are able to implicitly address bit errors. Through our model-aided DNN design as well as {training for a general Rayleigh-fading channel}, we are able to build a one-bit detector that doesn't need to be retrained for each new channel response. Finally, through our results we highlight the strength of the proposed approach, especially for the higher-order M-QAM constellations. 

Future directions of our work involve modifying this approach for mmWave channels, deployed for modern 5G networks. The presence of a larger power difference among multiple users, coupled with the correlated antenna measurements make one-bit detection for the mmWave channel case more challenging requiring further innovations in this DNN-aided one-bit detection paradigm.
%However, going forward we envision one-bit detection playing a vital role in wide-scale 5G communication network deployment. 

\section{{Acknowledgements}}
{
The authors would like to acknowledge support from PRP - Nautilus\footnote{{This work was supported in part by National Science Foundation (NSF) awards CNS-1730158, ACI-1540112, ACI-1541349, OAC-1826967, OAC-2112167, CNS-2100237, CNS-2120019, the University of California Office of the President, and the University of California San Diego's California Institute for Telecommunications and Information Technology/Qualcomm Institute. Thanks to CENIC for the 100Gbps networks.}} for access to high speed GPU resources for training and testing of our scripts. 
}

\bibliographystyle{IEEEtran}
% \mybibliography
\bibliography{Bib/references}

\end{document}